\documentclass[12pt]{article}
\pdfoutput=1

\usepackage[a4paper,text={16.8cm,22.4cm}]{geometry}

\usepackage{rotating}
\usepackage{cite}
\usepackage{amsmath}
\usepackage{amsfonts}
\usepackage{cancel}

\newcommand{\simge}{\hspace*{0.2em}\raisebox{0.5ex}{$>$}
     \hspace{-0.8em}\raisebox{-0.3em}{$\sim$}\hspace*{0.2em}}
\newcommand{\simle}{\hspace*{0.2em}\raisebox{0.5ex}{$<$}
     \hspace{-0.8em}\raisebox{-0.3em}{$\sim$}\hspace*{0.2em}}

\newcommand{\ep}{\epsilon}

\newcommand{\al}{\alpha}
\newcommand{\bt}{\beta}
\newcommand{\g}{\gamma}

\newcommand{\simu}{\sigma^{\mu\nu}}
\newcommand{\Fmu}{F_{\mu\nu}}
\newcommand{\Gmu}{G^a_{\mu\nu}}

\newcommand{\slashP}{\cancel{P}}

\newcommand{\slashT}{\cancel{T}}
\newcommand{\slashTsub}{\slash\hspace{-0.4em}T}
\newcommand{\slashCP}{\cancel{CP}}

\newcommand{\slashS}{\cancel{S}}

\newcommand{\qb}{\bar q}

\newcommand{\Fp}{F_\pi}

\newcommand{\mpi}{m_{\pi}}
\newcommand{\MQCD}{M_{\mathrm{QCD}}}

\newcommand{\Or}{\mathcal O}

\newcommand{\dslash}[1]{#1 \llap{/\kern-0.5pt}}
\newcommand{\Dslash}[1]{#1 \llap{/\kern+1.2pt}}
\newcommand{\DDslash}[1]{#1 \llap{/\kern+2.3pt}}
\newcommand{\dslashh}[1]{#1 \llap{/\kern+1pt}}

\newcommand{\boldtau}{\mbox{\boldmath $\tau$}}
\newcommand{\boldpi}{\mbox{\boldmath $\pi$}}

\def\slashchar#1{\setbox0=\hbox{$#1$}           
  \dimen0=\wd0                                    
  \setbox1=\hbox{/} \dimen1=\wd1                  
  \ifdim\dimen0>\dimen1                           
    \rlap{\hbox to \dimen0{\hfil/\hfil}}            
    #1                                             
  \else                                          
    \rlap{\hbox to \dimen1{\hfil$#1$\hfil}}        
    /                                           
 \fi}                                           %

\begin{document}

\begin{titlepage}

\begin{flushright}
  LAUR-15-23119
\end{flushright}

\markboth{E. Mereghetti and U. van Kolck}{Effective Field Theory and Time-Reversal Violation}

\vspace{0.2cm}
\begin{center}
\LARGE\bf

Effective Field Theory and
Time-Reversal Violation
in Light Nuclei

\end{center}

\vspace{0.2cm}

\vspace{0.2cm}
\begin{center}

{\large \bf E. Mereghetti$^{1}$} {\large and} {\large \bf U. van Kolck$^{2,3}$}

\vspace{0.6cm}
{\large  \it

$^{1}$Theoretical Division, Los Alamos National Laboratory, \\
Los Alamos, NM 87545, USA

$^{2}$Institut de Physique Nucl\'eaire, CNRS/IN2P3,
Universit\'e Paris Sud, \\ 91406 Orsay, France

$^{3}$Department of Physics, University of Arizona, \\
Tucson, AZ 85721, USA
}

\end{center}

\vspace{0.2cm}

\begin{center}

Accepted for publication in Ann. Rev. Nucl. Part. Sci. {\bf 65} (2015)

\end{center}

\vspace{0.2cm}

\begin{abstract}
Thanks to the unnaturally small value of
the QCD vacuum angle $\bar{\theta} \simle 10^{-10}$,
time--reversal violation ($\slashT$) offers 
a window into physics beyond the 
Standard Model (SM) of particle physics.
We review the effective-field-theory framework that
establishes a clean connection between $\slashT$ mechanisms,
which can be represented by 
higher-dimensional operators involving SM fields and symmetries,
and hadronic interactions,
which allow for controlled calculations of low-energy observables
involving strong interactions.
The chiral properties
of $\slashT$ mechanisms leads to a pattern 
that should be identifiable in measurements of
the electric dipole moments of the nucleon and light nuclei.
\end{abstract}



\end{titlepage}

\tableofcontents
\section{Introduction}

Violation of time reversal ($T$)  
is a fundamental asymmetry between past and future, 
the microscopic dynamics not 
being invariant under change in the sign of time.
In a Lorentz-invariant quantum field theory, where the product $CPT$ is 
conserved,
$T$ violation ($\slashT$) is equivalent to violation of 
the product of charge conjugation ($C$) ---the exchange between
particle and antiparticle--- and parity ($P$) ---the change
of sign in spatial coordinates.
$CP$ violation ($\slashCP$)
is one of the ingredients \cite{Sakharov:1967dj} needed to explain
why the visible universe seems to be made predominantly of matter,
without a significant fraction of antimatter \cite{Steigman:1976ev}.

The Standard Model (SM) \cite{Weinberg:1967tq,Salam:1968rm}
contains a source of $\slashCP$ and $\slashT$: the phase of 
the CKM matrix \cite{Kobayashi:1973fv},
which appears in observables through a combination 
$J\simeq 3\cdot 10^{-5}$ of matrix elements \cite{Jarlskog:1985ht}.
While this mechanism explains the violation observed in $K$ and $B$
decays \cite{Agashe:2014kda}, it
gives only very small contributions to 
quantities that do not involve flavor change between
initial and final states.
In particular, it is not sufficiently large to account for the 
observed matter-antimatter asymmetry \cite{Canetti:2012zc}. 

The ideal observables to probe new $\slashT$ interactions
are flavor-conserving observables such as 
$\slashT$ electromagnetic form factors (FFs), which
can be split into electric, magnetic, and toroidal,
depending on whether they interact with long-range electric,
long-range magnetic, or short-range electromagnetic fields.
Their $P$ and $T$ transformation properties are summarized in 
Table \ref{tab1}.
Of particular interest are permanent electric dipole moments (EDMs),
which require both $\slashP$ and $\slashT$.
Even with current technology they effectively probe very
small distances. For example,
the existing bound
on the neutron EDM, $|d_n| < 2.9 \cdot 10^{-13} e$ fm \cite{Baker:2006ts},
means that a charge imbalance, if any, effectively takes place at a distance
13 (or more) orders of magnitude smaller than the size of the neutron.
Other $\slashT$ multipoles, like magnetic (MQM) and toroidal (TQM)
quadrupole moments, are less accessible experimentally.

\begin{table}
\tabcolsep10pt
\label{tab1}
\begin{center}
\begin{tabular}{c|ccc}
{\bf Multipolarity} &{\bf Electric} &{\bf Magnetic} &{\bf Toroidal} \\
\hline
0 (monopole) &     $PT$    &    ---    &   --- \\
1 (dipole)   & $\slashP\slashT$ &   $PT$    & $\slashP T$  \\
2 (quadrupole) &     $PT$    &  $\slashP\slashT$ &  $P \slashT$  \\
$\cdots$ & $\cdots$ & $\cdots$ & $\cdots$ \\
\hline
\end{tabular}
\end{center}
\caption{Parity and time-reversal properties
of electromagnetic form factors according to multipolarity.
$S$ ($\slashS$) denotes that the symmetry $S$ is preserved (violated).
The pattern repeats as multipolarity increases.
A particle of spin $s$ has multipoles up to $2s$.
}
\end{table}

Searches are in progress around the world
for EDMs of the neutron and of (neutral) atoms and molecules, 
which are sensitive to the EDMs of the electron and nuclei, and their
$\slashT$ interactions. A new generation of experiments 
\cite{Kumar:2013qya}
promises to improve neutron EDM sensitivity by one or two orders of magnitude,
which is remarkable but
still above the expected CKM ``background'' at $\sim 10^{-19} e$ fm
(see Ref. \cite{Pospelov:2005pr} for an assessment 
and references to original papers, 
and Ref. \cite{Seng:2014lea} for a recent discussion).
Even more exciting is the groundbreaking proposal 
(see Refs. \cite{Semertzidis:2011qv,Pretz:2013us} for summaries and references)
that the EDMs 
of charged particles be investigated in specifically
designed storage rings, and not just as byproducts of other experiments
as for the muon \cite{Bennett:2008dy}.
We might see the deuteron EDM ($d_d$) probed at the level of
$\sim 10^{-16} e$ fm \cite{Pretz:2013us}, and similarly for 
the EDMs of the proton ($d_p$) and of the nucleus of helium-3, helion 
($d_h$).

The discovery of an EDM above the CKM background would be
a signal of long-sought new physics, but would, by itself,
leave us in the dark about its origins. 
{}From a theoretical perspective, it is important
to investigate the set of EDMs that would allow us to identify
the dominant source(s) of $\slashT$. 
Such an identification is likely to give clues
about physics beyond the SM (BSM) and the scale of this new physics,
which we denote $M_{\slashTsub}$.
The aim of this review
is to show that the framework of effective field theories (EFTs),
coupled to 
recent and not-too-distant-future progress in strong-interaction
physics, will allow us to carry out this identification
for the $\slashT$ operators that involve quarks and gluons.

There are good experimental and theoretical reasons to believe that the SM
is an EFT for processes involving
momenta $Q\sim M_{\textrm{EW}}\sim 100$ GeV.
In addition to the possible existence of new light or stable heavy particles,
new physics can be represented at the electroweak (EW) scale $M_{\textrm{EW}}$ by 
operators of canonical dimensions $d> 4$,
which involve known particles and are constrained by the SM symmetries,
namely Lorentz invariance and gauged
$SU(3)_c \times SU(2)_L\times U(1)_Y$.
One expects these operators to have strengths
${\cal O}(M_{\slashTsub}^{4-d})$, making the lowest-dimension operators
most significant.

The SM has other $d=4$
$\slashT$ operators, which involve the non-Abelian gauge bosons
and, despite being total derivatives, 
can contribute to $\slashCP$ observables due to topological effects
\cite{'tHooft:1976up}.
The operator involving gluons
could give rise to large EDMs of hadrons and nuclei,
but the neutron EDM bound already
significantly constrains its dimensionless
strength, the QCD vacuum angle $\bar{\theta}\simle 10^{-10}$.
This unnatural value is the famous ``strong $CP$ problem''.
The most promising solution is offered
by the Peccei-Quinn (PQ) mechanism  \cite{Peccei:1977hh},
where an additional approximate symmetry, $U(1)_{PQ}$,
is spontaneously broken generating a small $\bar{\theta}$ dynamically.
The corresponding pseudo-Goldstone boson,
the axion
\cite{Weinberg:1977ma,Wilczek:1977pj}, is a viable dark-matter candidate.
Whatever the mechanism may be, the smallness of $\bar{\theta}$
leaves room for higher-dimensional operators.

The sole $d=5$ interaction among known particles 
\cite{Weinberg:1979sa}
gives rise to neutrino masses and lepton-number violating processes, 
which are searched for with neutrinoless double-beta decay 
(see Ref. \cite{Bilenky:2014uka} for a recent review).
One can expect $\slashCP$ violation from phases in
the corresponding PMNS mixing matrix \cite{Bilenky:1980cx},
but it is unclear if the observed baryon-antibaryon asymmetry
can be generated through leptogenesis 
\cite{Fukugita:1986hr}, a mechanism based on the simplest
ultraviolet (UV) completion of the $d=5$ interaction.

The many $d=6$ interactions \cite{Buchmuller:1985jz} 
have been conveniently cataloged in Ref. \cite{Grzadkowski:2010es}.
Among those responsible for $\slashCP$ \cite{DeRujula:1990db},
hadronic and nuclear EDMs are most sensitive to
quark EDMs (qEDMs) and chromo-EDMs (qCEDMs),
the gluon chromo-EDM (gCEDM) \cite{Weinberg:1989dx},
and certain four-quark 
interactions 
\cite{RamseyMusolf:2006vr,Ng:2011ui}.
Nowadays the techniques exist 
---renormalization-group (RG) running down to the QCD scale $\MQCD\sim 1$ GeV,
lattice QCD (LQCD) for the calculation of low-energy constants (LECs),
nuclear EFTs to describe the dynamics at momenta $Q< \MQCD$
in terms of the LECs--- 
to connect these operators at the EW scale to light-nuclear EDMs. 

A frequent misconception is that nuclear-physics errors would obfuscate
any of the minute $\slashT$ effects we are interested in.
This presumption is obviously not correct for quantities like EDMs,
which vanish when $\slashT$ parameters vanish; in this case
errors affect only the proportionality factor, and as we are going to see
they should not affect many of the conclusions.
Light nuclei are special in this regard.
A key ingredient is the approximate chiral symmetry of QCD, and its breaking.
Not only does chiral symmetry provide the basis for a systematic expansion
of hadronic and nuclear observables in powers of 
$Q/\MQCD$ \cite{Bedaque:2002mn},
but it also acts as a ``filter'' to separate the effects of various
$\slashT$ sources. 
While all these sources generate 
$\slashT$ observables, they break chiral symmetry in different ways
and as a consequence produce different
patterns in the relative magnitudes for these observables.
Chiral symmetry was already an important aspect of the classic 
studies of the neutron EDM from $\bar{\theta}$ 
\cite{Baluni:1978rf,Crewther:1979pi},
and now it has been extended to other sources and more nucleons 
\cite{Mereghetti:2010tp,Maekawa:2011vs,deVries:2012ab,Bsaisou:2014oka}.
A single measurement (say, $d_n$) can always be attributed to 
any one source (say, a $\bar{\theta}$ of just the right, minute size),
but, as we review in Sections \ref{Sec4} and \ref{Sec5}, combined measurements of $d_n$,
$d_p$, $d_d$ and $d_h$ provide increasingly detailed information
on the $\slashT$ sources.
The further measurement of the triton EDM ($d_t$) would
allow as good a separation of underlying mechanisms as possible
in the strong-interacting sector at low energies, under the
assumption that lower-dimension operators are most important.
(Measurements of 
the deuteron MQM and TQM would also be valuable,
but seem impossible for the foreseeable future.)

All the main techniques needed for this analysis have experienced
significant progress recently. 
RG running down
to the QCD scale has long been used as a tool to investigate the
low-energy consequences of specific BSM models,
and a model-independent summary of these results has appeared
recently \cite{Dekens:2013zca}.
Computational advances are bringing LQCD to the forefront 
of hadronic and nuclear physics, and the time is approaching 
when EFTs will be able to use LQCD data, rather than experiment,
as input in the calculation of nuclear properties \cite{Barnea:2013uqa}.
Still, the study of $\slashT$ matrix elements is in its infancy,
particularly for $d>4$ operators,
and this 
constitutes the biggest gap in connecting nucleon and light-nuclear EDMs
to BSM $\slashT$ interactions.
Nuclear EFTs \cite{Bedaque:2002mn}
make it possible to approach hadronic and nuclear physics
incorporating SM symmetries,
and the $\slashT$ hadronic interactions involve, in lowest order,
six LECs.
Nuclear potentials inspired by Chiral EFT 
\cite{Machleidt:2011zz,Epelbaum:2012vx}, where pions and chiral
symmetry play a significant role, are
now the favorite starting point for
``{\it ab initio}'' nuclear-structure methods.
Although fully consistent calculations are not yet possible,
for light nuclei several tests suggest that errors
are relatively small and can be better quantified once
some subtleties in the RG of pion exchange are clarified 
\cite{Nogga:2005hy,Valderrama:2014vra}.
A significant future step would be to extend the framework summarized here 
to heavier nuclei, in order 
to enable a consistent analysis of atomic/molecular experiments
as well.

We limit ourselves here to a review of the techniques that
address $\slashT$ in the nucleon and light nuclei from a model-independent
perspective.
The relevant operators at
the quark/gluon level, including the RG to $\MQCD$,
are introduced in Section \ref{Sec2} and translated into hadronic
interactions in Section \ref{Sec3}.
The PQ mechanism is briefly summarized in Section \ref{Sec2.3}.
Sections \ref{Sec4} and \ref{Sec5} review calculations
of $\slashT$ electromagnetic observables for the nucleon
and light nuclei, respectively, starting
from the hadronic interactions.
An outlook is reserved for Section \ref{Sec6}. 
A much more comprehensive review ---which includes
implications of specific BSM models, calculations of QCD matrix elements with
various assumptions, calculations of heavy-nuclear
$\slashT$ quantities, a discussion of atomic/molecular EDMs,
and many more references to earlier work---
has appeared recently \cite{Engel:2013lsa}, and we refer the
reader to it for a more detailed look at how 
the program presented here relates to other efforts.

\section{$\slashT$ at the Quark-Gluon Level}
\label{Sec2}

We summarize here the most important $\slashT$ interactions among quarks,
gluons and photons. This background will allow the construction of
$\slashT$ hadronic interactions in Section \ref{Sec3}. 

\subsection{QCD $\bar\theta$ Term and Higher-Dimensional Operators}
\label{Sec2.1}

BSM physics can be described below its characteristic scale 
$M_{\slashTsub}>M_{\textrm{EW}}$
in terms of 
$SU(3)_c \times SU(2)_L\times U(1)_Y$ and Lorentz-symmetric operators.
The kinetic terms involving quarks interacting with
gluons $G_{\mu}^a$ ($a=1, ..., 8$)
and weak bosons $W_{\mu}^i$ ($i=1, 2, 3$) and $B_{\mu}$
with strengths $g_s$, $g$, and $g'$, respectively, 
are
\begin{equation}
\mathcal{L}_{T}^{(4)}=  \bar q_L i \slashchar{D} q_L
+ \bar u_R i \slashchar{D} u_R
+ \bar d_R i \slashchar{D} d_R
-\frac{1}{4} 
\left(G^a_{\mu\nu}G^{a\mu\nu} + W^i_{\mu\nu}W^{i\mu\nu} + B_{\mu\nu}B^{\mu\nu}\right) , 
\label{eq:2.00}
\end{equation}
where $q_L$ is a doublet of left-handed quarks;
$u_R$ and $d_R$ are right-handed up- and down-type quarks;
$D_\mu=\partial_{\mu} - i g_s G_{\mu}^at_a
-ig W_{\mu}^i t_i  -i g' B_{\mu} Y$ is the gauge covariant derivative
with $t_a=\lambda_a/2$ ($\lambda_a$ are the Gell-Mann matrices),
$t_i=\tau_i/2, 0,0$ ($\tau_i$ are the Pauli matrices)
and $Y=1/6,2/3,-1/3$ for $q_L$, $u_R$ and $d_R$, respectively;
and 
$G_{\mu\nu}^a=\partial_\mu G_{\nu}^a-\partial_\nu G_{\mu}^a-g_sf^{abc}G_{\mu}^bG_{\nu}^c$, 
$W^i_{\mu \nu}=\partial_\mu W_{\nu}^i-\partial_\nu W_{\mu}^i
-g\epsilon^{ijk}W_{\mu}^j W_{\nu}^k$,
and 
$B_{\mu \nu}=\partial_\mu B_{\nu}-\partial_\nu B_{\mu}$
are, respectively, the 
$SU(3)_c$, $SU(2)_L$, and $U(1)_Y$ field strengths
with structure constants $f^{abc}$ for $SU(3)_c$ and $\ep^{ijk}$
for $SU(2)_L$.
For simplicity we omit indices that run through
the three generations, which are summed over.

The $d=4$ terms related to $\slashT$ are 
the Yukawa couplings of the quarks and the topological 
$\theta$ term \cite{'tHooft:1976up},
\begin{equation}
\mathcal L_{\slashTsub}^{(4)}  = 
-\left( \bar q_L  Y^{u} \tilde \varphi \, u_R 
+ \bar q_L Y^{d} \varphi  \, d_R \right) +\mathrm{H.c.}
- \theta\frac{g^2_s}{64\pi^2}\, \ep^{\mu\nu\al\bt} \, G^a_{\mu\nu}G^a_{\al\bt} \ ,
\label{eq:2.01}
\end{equation}
where
$\varphi$ is the Higgs doublet and
$\tilde{\varphi}^I=\epsilon^{IJ}\varphi^{J*}$.
(Here $\ep^{01}=\ep^{012}=\ep^{0123}=+1$.)
The Yukawa couplings $Y^{u,d}$ form $3 \times 3$ complex matrices in 
flavor space. Relative phases lead to $\slashCP$ in the CKM matrix 
\cite{Kobayashi:1973fv} and,
as described below, there is an interplay between the overall phase
and the vacuum angle $0\le \theta <2\pi$.
A term analogous to $\theta$ for weak gauge bosons 
gives negligible contributions at low energies.

Because of the smallness of $d=4$ $\slashT$,
we follow Refs. \cite{deVries:2012ab,Bsaisou:2014oka} 
and consider the $d=6$ terms \cite{Grzadkowski:2010es},
\begin{eqnarray}
\mathcal L_{\slashTsub}^{(6)} &=& 
-2 \frac{\varphi^{\dagger} \varphi}{v^2}  
\left[
\left(\bar q_L  Y^{\prime\, u} \tilde \varphi u_R 
+\bar q_L Y^{\prime\, d} \varphi d_R\right) +\mathrm{H.c.}
+\theta' \frac{g_s^2 }{64\pi^2} 
\, \ep^{\mu\nu\al\bt}\, G^a_{\mu\nu}G^a_{\al\bt} \right]
\nonumber\\
&&
- \frac{1}{\sqrt{2}}\bar q_L \sigma^{\mu \nu} 
\left(g_s \tilde\Gamma^{u} t_a G^a_{\mu \nu} 
+ g \Gamma^u_{W} \tau_i W^i_{\mu \nu}  
+ g^{\prime}\Gamma^u_{B} B_{\mu \nu} \right) 
\tilde\varphi \, u_R  
+ \mathrm{H. c.}
\nonumber \\ 
&& 
- \frac{1}{\sqrt{2}} \bar q_L \sigma^{\mu \nu} 
\left( g_s \tilde\Gamma^{d} t_a G^a_{\mu \nu} 
+ g \Gamma^{d}_W \tau_i W^i_{\mu \nu} 
+ g^{\prime} \Gamma^{d}_B  B_{\mu \nu} 
\right) \varphi \,   d_R 
+ \mathrm{H. c.}
\nonumber\\
&&
+\frac{d_{W}}{6}g_s f^{a b c} \ep^{\mu \nu \alpha \beta} 
G^a_{\alpha \beta} G_{\mu \rho}^{b} G^{c\, \rho}_{\nu}
+g_s^2 \, \bar u_R \Xi_1 \gamma^\mu d_R  \;
\tilde \varphi^\dagger iD_\mu \varphi
+ \mathrm{H.c.} 
\nonumber \\ 
&& 
+g_s^2 \, \ep^{JK} \left[ 
\Sigma_1 \;\bar q^J_L u_R \;  \bar q_L^K d_R
+ \Sigma_8 \; \bar q^J_L t_a u_R \; \bar q_L^K t_a d_R\right]
+ \textrm{H.c.},
\label{eq:2.02}
\end{eqnarray}
where $v\simeq 246$ GeV is the Higgs vacuum expectation value (vev).
Here:

\begin{itemize}

\item 
The $3 \times 3$ complex matrices
$Y^{\prime\, u,d}$ and the angle $\theta'$ correct,
after EW symmetry breaking,
Yukawa couplings and $\theta$ term.
Terms linear in the Higgs field are 
of more phenomenological interest, as they 
either modify the Yukawa couplings of the Higgs to quarks
introducing, in general, flavor-changing effects,
or affect the gluon-fusion production mechanism  at the LHC 
\cite{Manohar:2006gz,Alonso:2013hga,Brod:2013cka,Kagan:2014ila}.

\item 
The coefficients $\tilde{\Gamma}^{u,d}$ and $\Gamma_{B,W}^{u,d}$ are 
$3 \times 3$ matrices in flavor space. 
The nondiagonal entries contribute to flavor-changing 
currents and play an important role in flavor physics 
\cite{Isidori:2011qw,Sala:2013osa}.
The diagonal components determine   
the quark electric (qEDM) and chromoelectric (qCEDM)
dipole moments, as well as the quark weak EDM, which 
plays only a minor role at low energy.

\item 
The parameter $d_W$
of the Weinberg three-gluon operator \cite{Weinberg:1989dx}
can be thought of as the
gluon chromoelectric dipole moment (gCEDM).
Similar terms involving weak gauge bosons (for example,
the $W^{\pm}$ boson's EDM and MQM \cite{Hagiwara:1986vm})
can be relevant at colliders, but are very small at energies 
below $M_{\textrm{EW}}$.

\item 
The $\Sigma_{1,8}$ 
are complex four-index tensors 
in flavor space. 
If one considers only quarks of the first generation, there are two 
$\slashCP$ four-quark operators that respect $SU(2)_L$ 
\cite{RamseyMusolf:2006vr}, which we refer to as PS4QOs due to
their pseudoscalar nature.
Additional four-quark operators can be constructed involving quarks of 
different generations. 

\item 
The complex $3 \times 3$ matrix $\Xi_1$ couples $W^\pm$ bosons to the
right-handed quark current. It leads below $M_{\textrm{EW}}$
to additional $\slashCP$ four-quark operators \cite{Ng:2011ui},
which couple left- and right-handed quarks and we call LR4QOs.
It also causes $\slashCP$ in nuclear $\beta$ decay when $W^{\pm}$ connect 
to the left-handed lepton current \cite{Ng:2011ui}.

\end{itemize}
The $d=6$ coefficients 
depend on the spectrum and 
$\slashCP$ parameters of the BSM model of choice,
and can be determined by matching just
below the scale $M_{\slashTsub}$.
While a detailed study 
is beyond the scope of this review (see instead Ref. \cite{Engel:2013lsa}), 
from our brief 
discussion it is apparent that there is a rich interplay among the 
constraints on the couplings in Eq. \eqref{eq:2.02}
that can be extracted from collider, flavor, and low-energy precision 
experiments.

Below the scale $M_{\textrm{EW}}$
the breaking of $SU(3)_c \times SU(2)_L\times U(1)_Y$ to 
$SU(3)_c \times U(1)_{em}$ is important and generates masses
for fermions and weak gauge bosons.
For studying low-energy observables like EDMs, the Lagrangian 
\eqref{eq:2.02} at the 
EW scale needs to be matched onto a theory with only light quarks, 
gluons and the photon $A_\mu= \sin \theta_W W^3_\mu+ \cos \theta_W B_\mu$,
where $g \sin \theta_W = g' \cos \theta_W=-e<0$.
For nuclear physics applications we can limit ourselves to two
light flavors, and at 
$M_{\textrm{QCD}}$
the kinetic terms can be written as
\begin{equation}
\mathcal{L}_{T}^{(4)}=  \bar q_L i \slashchar{D} q_L
+ \bar q_R i \slashchar{D} q_R
-\frac{1}{4} G^a_{\mu\nu}G^{a\mu\nu} 
-\frac{1}{4} F_{\mu \nu} F^{\mu \nu} , 
\label{eq:2.001}
\end{equation}
where $q = (u \; d)^T$,
$D_{\mu} = \partial_{\mu} -i g G^a_{\mu}t_a -i e A_{\mu} Q$ is the 
$SU(3)_c \times U(1)_{em}$ covariant derivative
in terms of the charge matrix $Q = 1/6+\tau_3/2$,
and $F_{\mu \nu}=\partial_\mu A_\nu-\partial_\nu A_\mu$ is the
$U(1)_{em}$ field strength.
Some of the issues associated with
strangeness are discussed in Section \ref{Sec3.3}.

$\slashCP$ from the QCD $\theta$ term 
is intimately related to the quark masses.  
All the phases of the quark mass matrix can be eliminated through non-anomalous
$SU(2)$ 
vector and axial rotations, except for a common phase $\rho$, leaving
\begin{equation}
\mathcal{L}_{\slashTsub}^{(4)}=  
- \left(e^{i \rho} \qb_L  M q_R + e^{-i \rho} \qb_R  M  q_L\right)
- \theta\frac{g^2_s}{64\pi^2} \ep^{\mu\nu\al\bt} \, G^a_{\mu\nu}G^a_{\al\bt} ,
\label{eq:2.1}
\end{equation}
where $M= \bar m (1- \varepsilon \tau_3)$
is the diagonal quark mass matrix 
with $\bar m$ (the average light-quark mass)
and $\varepsilon$ (the relative light-quark mass splitting) 
real parameters. 
The phase $\rho$
and the $\theta$ angle are not independent. 
By performing an anomalous axial $U(1)_A$ rotation, all 
$\slashCP$
can be rotated into the $\theta$ term or into the complex mass term,  
and physical observables depend only on the combination 
$\bar\theta = \theta + 2\rho$.
For our discussion in Chiral EFT,
it is 
convenient to eliminate the $\theta$ term in favor of a complex quark mass.
The additional $SU(2)$
approximate symmetry of the
Lagrangian can be exploited to align the 
vacuum 
in the presence of $\slashCP$
to the usual QCD vacuum \cite{Baluni:1978rf,Crewther:1979pi}.
From a low-energy point of view, vacuum alignment is equivalent to setting 
to zero the coupling of the neutral pion 
to the vacuum, 
at lowest order in 
the Chiral EFT expansion \cite{Mereghetti:2010tp}. 
The Lagrangian 
becomes 
\begin{equation}
\mathcal{L}_{\slashTsub}^{(4)} = 
-  \bar m \, r(\bar\theta) \, \bar q q 
+\bar m \, r^{-1}(\bar\theta)
\; \bar q \left(\varepsilon \, \tau_3   
 + \frac{1-\varepsilon^2}{2}\sin\bar\theta\; i \gamma_5 
\right) q,
\label{eq:2.2}
\end{equation} 
where $r(x)$ is an even function of $x$, 
\begin{equation}
r(x) = \sqrt{\frac{1 + \varepsilon^2 \tan^2 \frac{x}{2}}{1+\tan^2 \frac{x}{2}}}
= 1 - (1 - \varepsilon^2)\frac{x^2}{8} +{\cal O}(x^4) .
\label{eq:2.3}
\end{equation}
The $\mathcal O(\bar\theta^2)$ terms become important 
in the PQ mechanism (Section \ref{Sec2.3}).
The last term in Eq. \eqref{eq:2.2} is approximately
linear in $\bar\theta$, and responsible for $\slashCP$.

In the absence of flavor change, the $d=6$ $\slashT$ operators that receive 
tree-level contributions and are not suppressed by powers of $M_{\textrm{EW}}$ are 
\cite{deVries:2012ab}
\begin{eqnarray}
\mathcal{L}_{\slashTsub}^{(6)}&=&
-\frac{\bar m}{2} \qb \left(d_0 Q 
+\frac{d_3 }{2} \left\{ Q, \tau_3 \right\}\right)i \simu \g_5 q \; e \Fmu 
-\frac{\bar m}{2} \qb \left(\tilde{d}_0+\tilde{d}_3 \tau_3\right)
i \simu\g_5 t_a q \; g_s \Gmu
\nonumber\\
&&
+ \frac{d_{W}}{6} g_s f^{a b c} \ep^{\mu \nu \alpha \beta} 
G^a_{\alpha \beta} G_{\mu \rho}^{b} G^{c\, \rho}_{\nu} 
\nonumber\\
&&
+ \frac{g_s^2}{4} 
\left[\textrm{Im}{\Sigma_1} \left( \bar q q\, \bar q i \gamma_5 q 
- \bar q \boldtau q \cdot \bar q \boldtau i \gamma_5 q \right)
+ \textrm{Im}{\Sigma_8}\left( \bar q t_a q\, \bar q i \gamma_5 t_a q 
- \bar q \boldtau t_a q \cdot \bar q \boldtau i \gamma_5 t_a q 
\right)\right]
\nonumber\\
&& 
+ \frac{g_s^2}{4} \ep^{3ij}
\left(\textrm{Im}\Xi_1 \, \bar q \tau_i \gamma^{\mu}q \, 
\bar q \tau_j \gamma_{\mu} \gamma_5 q  
+ \textrm{Im}\Xi_8 \, \bar q \tau_i \gamma^{\mu} t_a q 
\, \bar q \tau_j  \gamma_{\mu} \gamma_5 t_a q\right).
\label{eq:2.9}
\end{eqnarray}
The tree-level matching of Eq. \eqref{eq:2.9} to Eq. \eqref{eq:2.02}
is, for most operators, trivial. 
The 
qEDMs and qCEDMs are 
\begin{equation}
\bar m \, d_{0, 3} = 
- \frac{3v}{4} \, \textrm{Im}\left[ 
\left(\Gamma^u_B + \Gamma^u_W\right)_{11}
\mp 2 \left(\Gamma^d_B -\Gamma_W^d\right)_{11}\right],
\quad
\bar m \,\tilde{d}_{0, 3} =
\frac{v}{2} \, \textrm{Im}
\left(\tilde\Gamma^u \pm \tilde\Gamma^d \right)_{ 11 },
\label{eq:2.15} 
\end{equation}
where the indices refer to generations.
In most models $(\Gamma_{B,W}^{u,d})_{11}$
and $(\tilde \Gamma)_{11}$ are proportional to the light-quark Yukawa couplings,
thus canceling the light-quark mass on the left-hand-side  of Eq. \eqref{eq:2.15}.
The gCEDM gets tree-level contributions only from itself, 
and similarly for PS4QOs.
The imaginary part of the right-handed current in Eq. \eqref{eq:2.02} 
contributes to the $SU(2)_L$-breaking LR4QOs. Thus, at tree level,
\begin{equation}
\textrm{Im} \Sigma_{1,8} = \left(\textrm{Im} \Sigma_{1,8} \right)_{11 11} , 
\qquad 
\textrm{Im} \Xi_{1} =  V_{ud} \left(\textrm{Im} \Xi_{1}\right)_{11}, 
\qquad 
\textrm{Im}\Xi_8 =  0,
\label{eq:2.16}
\end{equation}
where $V_{ud}\simeq 0.97$ is the up-down CKM element. 
$\textrm{Im}\Xi_8$ is, however,
generated by the QCD evolution,
as we discuss next.
Note that all interactions in Eq. \eqref{eq:2.9} are also $\slashP$;
$P\slashT$ interactions are effectively of higher order
and expected to produce even smaller effects.

\subsection{Sizes and Runnings of $\slashT$ Couplings}
\label{Sec2.2}

As the scale $\mu$ of interest decreases,
one has to evolve the $\slashCP$ coefficients. 
The $\bar \theta$ term is not multiplicatively renormalized
because $\bar \theta$ is periodic, although
it can mix with the divergence of the axial current \cite{Espriu:1982bw}.
The operators in Eq. \eqref{eq:2.02} 
run from just below $M_{\slashTsub}$ to $M_{\textrm{EW}}$
and the operators in Eq. \eqref{eq:2.9} run 
from just below $M_{\textrm{EW}}$ to $\MQCD$. 
Much of the literature concerns specific BSM models,
where particular operators are singled out. 
A model-independent analysis to one loop, 
focusing on flavor-conserving interactions of the first generation,
was carried out in Ref. \cite{Dekens:2013zca}, and
the more general situation is under study \cite{Dekens:2015}.

For  $M_{\slashTsub}\sim$ a few TeV,
the dominant effects in the evolution of coefficients 
are due to the strong interaction,
and can be written in terms
of the numbers of colors $N_c$ and flavors $n_f$, 
the Casimir $C_F=(N_c^2-1)/2N_c$,
and the beta function for $g_s=\sqrt{4\pi\alpha_s}$,
$\beta_0 = (11 N_c - 2n_f)/3$. 
The EW running for some of the operators has also been considered 
in Ref. \cite{Dekens:2013zca}, where earlier references can be found.
The collection $\vec{C}= (\vec{C}_1, \vec{C}_2, \vec{C}_3)^T$
of $d=6$ operators obeys the RG equation 
with a matrix $\gamma$ of anomalous dimensions,
\begin{equation}
\frac{d \vec{C}(\mu)}{d \ln \mu} = \gamma \vec C(\mu), 
\qquad
\gamma = \frac{\alpha_s}{4\pi} 
\left( \begin{array}{c c c c c}
\gamma_{\rm{dip}} & \gamma_{\rm{mix}} & \gamma_{13} \\
0              & \gamma_{\rm{PS}}  & 0           \\
0     & 0               & \gamma_{33} \\
\end{array} \right).
\label{eq:2.10}
\end{equation}
Here $\vec{C}_1= (d_0, d_3,  \tilde{d}_0, \tilde d_3, d_W)^T$ and
$\vec{C}_2= (\textrm{Im} \Sigma_1, \textrm{Im} \Sigma_8)^T$.
The renormalization and mixing 
of qEDMs, qCEDMs and gCEDM is described by 
\cite{Braaten:1990gq,Dekens:2013zca}
\begin{equation}
\gamma_{\rm{dip}} = \left(\begin{array}{c c c c c}
8 C_F & 0     & - 8 C_F         & 0        & 0 \\
0     & 8 C_F &      0          & - 8 C_F  & 0 \\
0     & 0     & 16 C_F - 4 N_c  & 0        & 2 N_c \\
0     & 0     &     0           &16 C_F - 4 N_c  & - 2\varepsilon N_c  \\
0     &  0    &     0           & 0              & \beta_0 + N_c + 2 n_f \\
\end{array}\right),
\label{eq:2.11}
\end{equation}
while 
\begin{eqnarray}
 \gamma_{\rm{PS}}  &=& 2 \left( \begin{array}{c c} 
\beta_0 -2 (3 + 4/N_c) C_F\;\; & 
-2(1+1/N_c)^2(N_c -2) C_F 
\\
4 (1+ 2/N_c)\;\; &  \beta_0+2(C_F -1- 2/N_c^2) \\
\end{array} \right),
\label{eq:2.12}
\\
\gamma_{\rm{mix}} &=& \frac{1}{2} 
\left( \begin{array}{c c c c c } 
-5 + 3 \varepsilon  & 3 - 5\varepsilon  & -4 & - 4 \varepsilon & 0 \\
(-5 + 3\varepsilon) C_F \; &  (3 - 5\varepsilon) C_F \;
& 2(N_c-2C_F) \; & 2\varepsilon (N_c-2C_F) \; & 0  \\
\end{array}
\right)^T,
\label{eq:2.13}
\end{eqnarray}
contain, respectively, the anomalous dimensions of 
singlet and octet PS4QOs
\cite{An:2009zh,Hisano:2012cc,Dekens:2013zca},
and the mixings of dipoles with PS4QOs
\cite{Hisano:2012cc,Dekens:2013zca}.
The first 
four columns of $\gamma_{\rm{dip}}$ are known to 
two loops \cite{Degrassi:2005zd}.

Above $M_{\textrm{EW}}$, $\vec{C}_3= (g_s^2\textrm{Im} \Xi_1,\theta^{\prime}, 
\textrm{Im} Y^{\prime\, u}, \textrm{Im} Y^{\prime\, d})^T$.
The dipoles mix only with the gluon-Higgs operator $\theta^{\prime}$,
whose diagonal entry vanishes
\cite{Grojean:2013kd,Dekens:2013zca},
the right-handed current operator $\textrm{Im} \Xi_1$
does not require renormalization \cite{Dekens:2013zca},
and the quark-Higgs couplings $\textrm{Im} Y^{\prime\, u,d}$
renormalize multiplicatively as the quark masses \cite{Dekens:2013zca},
so that
\begin{equation}
\gamma_{13} = -  \frac{1}{2\pi^2 v^2}
\left(\vec{0}\;\; \left(0 \; \; 0\;\; 1\; \; \varepsilon \;\; 0\right)^T 
\;\; \vec{0} \;\; \vec{0}\right),
\qquad
\gamma_{33} = -6 C_F\left(\begin{array}{c c c c}
0     & 0     & 0  & 0 \\
0     & 0     & 1  & 0\\
0     & 0     & 0  & 1 \\
\end{array}\right).
\label{eq:2.12b}
\end{equation}
In these expressions we assumed that the qCEDM is proportional to the
quark mass, and neglected terms with higher powers of quark masses.

Below $M_{\textrm{EW}}$, 
the quark-Higgs and gluon-Higgs operators 
disappear from the operator basis, and two new operators, 
the LR4QOs,
appear: now $\vec{C}_3= (\textrm{Im} \Xi_1, \textrm{Im} \Xi_8)^T$.
These operators do not mix with the remaining operators and
we have \cite{An:2009zh,Hisano:2012cc,Dekens:2013zca}
\begin{equation}
\gamma_{13} = \left(\vec{0}\;\; \vec{0}\right),
\qquad
\gamma_{33}  = 2 \left( \begin{array}{c c} 
\beta_0\quad &  - 3 C_F/N_c \\
-6 \quad     &  \beta_0 - 3N_c(1 - 2/N_c^2) \\
\end{array} \right).
\label{eq:2.12LR}
\end{equation}

Besides the tree-level matching to coefficients of the Lagrangian 
\eqref{eq:2.9}, other tree-level contributions exist. 
For example, the quark-Higgs operator
contributes to chiral-symmetry-breaking four-quark operators.
However, these contributions are suppressed by additional powers 
of the EW scale, and are effectively of higher dimension ---we refer to
Ref. \cite{Dekens:2013zca} for a discussion.
Most loop corrections involve powers of $\alpha_{\textrm{em}}$, 
which for simplicity we ignore.
The gCEDM receives threshold corrections from the top CEDM, 
and, at lower energy, from the bottom and charm CEDMs \cite{Braaten:1990gq}. 
At one loop, these contributions are finite, and, with our conventions,
amount to a shift 
\begin{equation}
\delta d_W (m_Q)= - \frac{\alpha_s(m_Q)}{8 \pi} \tilde{d}_Q(m_Q) 
\qquad (Q=t,c,b).
\label{eq:2.17}
\end{equation}
In the absence of the PQ mechanism, 
heavy quark CEDMs are also constrained 
by the large radiative corrections they induce in  $\bar\theta$
\cite{Chang:1991hz}. 
When the Higgs is integrated out, $\bar\theta^\prime$ generates an 
$\mathcal O(\alpha_s)$ threshold correction to the qCEDM, 
whose effect is however smaller than the qCEDM induced by running.
Other important threshold corrections involve the top Yukawa coupling, 
and arise at two loops, through Barr-Zee type diagrams \cite{Barr:1990vd}.

\begin{figure}[t]\center
\includegraphics[width=10cm]{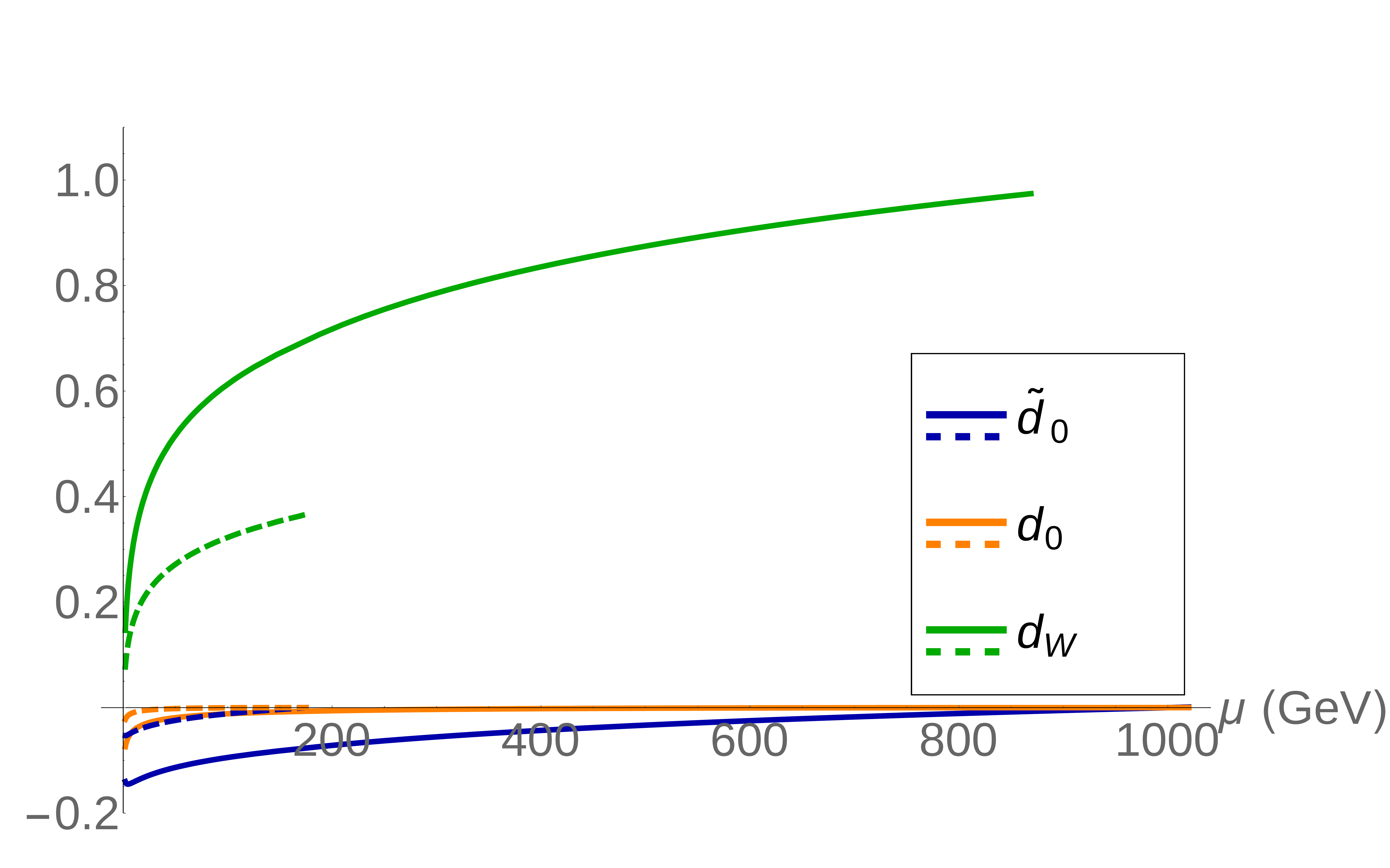}
\caption{Examples of the RG evolution of 
$d=6$ $\slashCP$
operators: qCEDM ${\tilde d}_0$, qEDM $d_0$, and gCEDM $d_W$
(in units of $M_{\slashTsub}^{-2}$)
as functions of the scale $\mu$ (in GeV).
The only nonvanishing operator at $M_{\slashTsub}\sim$ 1 TeV
is taken to be the gCEDM (continuous lines) or the top CEDM (dashed lines).
For visibility, the qCEDM, qEDM and gCEDM originating
in the tCEDM are multiplied by 100.}
\label{Fig1}
\end{figure}

The overall effect of the RG is to modify the coefficients by
factors which are typically of ${\cal O}(1)$
(except for $\textrm{Im} \Sigma_1$ which can get enhanced
by almost an order of magnitude \cite{Dekens:2013zca}).
In Fig. \ref{Fig1}  we illustrate effects of RG evolution.
We consider cases where at the scale  
$M_{\slashTsub}$ only the gCEDM  (continuous lines) or the top CEDM (dashed lines) 
exists.
For the sake of illustration, we took 
$\tilde{d}_{t}(M_{\slashTsub}) = - 100\, d_W(M_{\slashTsub})$.
In the  first case, we see that RG evolution reduces $d_W$  
to about $20\%$ of its original value, 
while generating a qCEDM and, to a lesser extent, a qEDM. 
While we plotted only the isoscalar components,
also the isovector qCEDM and qEDM are generated, 
in the proportion $\tilde d_3 / \tilde d_0 = d_3/d_0 = \varepsilon \sim 1/3$.
The tCEDM  contributes to the gCEDM at the top threshold,
Eq. \eqref{eq:2.17}, 
and generates light-quark qEDM and qCEDM through RG evolution.   
Although the induced gCEDM is a factor of 1000 smaller than 
$\tilde d_t$, the constraint from the neutron EDM is still 
about two orders of magnitude stronger than the direct bound
from $t\bar t$ production at the LHC \cite{Kamenik:2011dk}.

The main outcome of this analysis is that at $\mu\sim \MQCD$
the best organizational principle is given not by canonical dimension, 
but by the effective
dimension inherited from the SM Lagrangian, Eq. \eqref{eq:2.02}.
Thus all the coefficients in Eq. \eqref{eq:2.9}, including
the q(C)EDMs, have sizes
consistent with effective dimension six \cite{deVries:2012ab}:
\begin{equation}
d_{i} =\Or\!\left(\frac{\delta_{i}}{M^2_{\slashTsub}}\right),
\,
\tilde d_{i} =
\Or\!\left(\frac{\tilde\delta_{i}}{M^2_{\slashTsub}}\right),
\,
d_W = \Or\!\left(\frac{w}{M^2_{\slashTsub}}\right),
\,
\mathrm{Im}\Sigma_{a} = \Or\!\left(\frac{\sigma_{a}}{M^2_{\slashTsub}}
\right),
\,
\mathrm{Im}\Xi_{a} = \Or\!\left(\frac{\xi}{M^2_{\slashTsub}}\right),
\label{scalingofdim6}
\end{equation}
where $\delta_{0,3}$, $\tilde\delta_{0,3}$, $w$, $\sigma_{1,8}$, and $\xi$
are eight real parameters encoding details of the BSM physics.
Naive dimensional analysis (NDA) \cite{Manohar:1983md,Georgi:1986kr}
suggests that these parameters are ${\cal O}(1)$.
In contrast, six other $\slashCP$ four-quark operators 
which are invariant under $SU(3) \times U(1)_{em}$ 
\cite{An:2009zh,Hisano:2012cc} originate from either
higher-dimensional operators at $M_{\textrm{EW}}$
or higher-order electroweak effects, and are suppressed with
respect to the PS4QOs and LR4QOs kept in Eq. \eqref{eq:2.9}
\cite{deVries:2012ab,Dekens:2013zca}.

\section{$\slashT$ at the Hadronic Level}
\label{Sec3}

The $\slashT$ interactions among quarks and gluons discussed in 
Section \ref{Sec2} translate, at low energies, into $\slashT$ interactions 
involving the lightest mesons and baryons, which we sketch in this section.

\subsection{Chiral Symmetry and Low-Energy Interactions}
\label{Sec3.1}

At a momentum comparable to the pion mass, $Q\sim m_\pi\ll \MQCD$, 
the implications of the Lagrangian in Eqs. 
\eqref{eq:2.2}  and \eqref{eq:2.9} for the interactions among pions
and nucleons are described by Chiral Perturbation Theory ($\chi$PT) 
\cite{Weinberg:1978kz,Gasser:1983yg,Gasser:1984gg}
and its extension to arbitrary number of nucleons, Chiral EFT 
\cite{Weinberg:1990rz,Weinberg:1991um,Ordonez:1992xp}.
The special role of the pion is a consequence of the 
invariance of Eq. \eqref{eq:2.001} (for $e=0$)
under the chiral symmetry $SU(2)_L\times SU(2)_R\sim SO(4)$, 
and its spontaneous breaking to the isospin subgroup $SU(2)_V \sim SO(3)$. 
Pions emerge as Goldstone bosons whose interactions are 
proportional to their momenta, 
which guarantees that low-momentum observables can be 
computed in a perturbative expansion in powers of $Q/M_{\textrm{QCD}}$.
Quark masses and other interactions 
explicitly break chiral symmetry,
give pions masses, and induce non-derivative pion couplings,
but the breaking is small and can be incorporated in the expansion.
The chiral Lagrangian contains an 
infinite number of operators which 
can be grouped using a ``chiral index'' $\Delta=d+f/2-2\ge 0$ that
increases with the number $d$ of derivatives and chiral-symmetry
breaking parameters, and the number $f$ of fermions.
Assuming NDA,
the prediction of any observable at a given accuracy in $Q/M_{\textrm{QCD}}$ 
requires the consideration of only a finite number of operators, 
up to a certain $\Delta$.

Chiral symmetry is realized nonlinearly in the chiral Lagrangian
\cite{Weinberg:1968de,Coleman:1969sm,Callan:1969sn}, whose construction via
chiral-covariant objects is well known \cite{Weinberg:1996kr}.
The choice of fields is arbitrary, and for definiteness we employ 
a stereographic parametrization of the isospin-triplet $\boldpi$, 
and an isospinor $N=(p\, n)^T$ for the nucleon.
In this case the 
chiral-covariant derivatives take the form
$D_{\mu}\pi_i = D^{-1} (\partial_{\mu} \delta_{ij}+ e A_\mu \ep_{3ij})\pi_j$
and 
${\mathcal D}_\mu N= 
[\partial_\mu+i\boldtau\cdot\boldpi\times D_{\mu}\boldpi/F_\pi^2
+ieA_\mu (1+\tau_3)/2]N$,
with $F_{\pi} \simeq 186$ MeV the pion decay constant\footnote{ Note that the pion decay constant is frequently defined
in the literature as $f_\pi = F_\pi/2$ or $f_\pi = F_\pi/\sqrt{2}$. } and
$D = 1 + \boldpi^2/F_{\pi}^2$.
Isospin-invariant objects are automatically
chiral invariant;
chiral-variant interactions are built with the 
$SO(4)$ transformation properties of the corresponding quark-gluon 
interactions.
For example, electromagnetic interactions in Eq. \eqref{eq:2.001} break chiral
symmetry as an antisymmetric $SO(4)$ tensor \cite{vanKolck:1995cb}.
Because $m_N\sim \MQCD$, Chiral EFT is well defined only for non-relativistic
nucleons. The $Q/m_N$ expansion can be made consistent with
the $Q/\MQCD$ expansion with heavy nucleon fields
\cite{Jenkins:1990jv},
when the nucleon mass $m_N$ is removed from propagators
and the Dirac structure simplifies to 
the nucleon velocity $v^{\mu}$ and spin $S^{\mu}$.
Resummations of the $Q/m_N$ expansion are currently popular but they
do not decrease the overall theoretical error.
Chiral EFT can be extended to the Delta-isobar region 
using explicit fields for the Delta \cite{Pascalutsa:2002pi,Long:2009wq} 
and the Roper \cite{Long:2011rt}, but
most $\slashT$ applications have so far been limited to nucleons.
Chiral EFT reproduces the (B)SM $S$ matrix because 
it includes all operators consistent with RG invariance
and the symmetries of quark/gluon interactions at $\mu \sim \MQCD$.

Equation \eqref{eq:2.001} is represented
at lowest chiral index as
\begin{eqnarray}
\mathcal L^{(0)}_T &=& -\frac{1}{4} F_{\mu \nu} F^{\mu \nu} 
+ \frac{1}{2} D_{\mu} \boldpi \cdot D^{\mu} \boldpi 
+ \bar N \left( i v\cdot \mathcal D 
- \frac{2 g_A}{F_{\pi}} S^{\mu} \boldtau \cdot D_{\mu}\boldpi \right) N
\nonumber\\
&&+ C_{S}\bar N N\bar N N + C_{V}\bar N \boldtau N\cdot \bar N \boldtau N , 
\label{eq:3.1}
\end{eqnarray}
where 
$g_A ={\cal O}(1)\simeq 1.27$ is the pion-nucleon axial coupling
and $C_{S,V}$ 
are 
LECs related to the two nucleon-nucleon scattering lengths.
Higher-index interactions are constructed with further derivatives
and nucleon fields.
We assign a chiral index 3 to $\alpha_{\textrm{em}}/4\pi$,
powers of which purely hadronic operators 
generated by the integration of hard photons 
---{\it e.g.}, the electromagnetic pion mass splitting 
$\hat{\delta} m_\pi^2={\cal O}(\alpha_{\textrm{em}}\MQCD/4\pi)
\simeq 1260$ MeV$^2$ \cite{Agashe:2014kda}---
are proportional to. 

The properties under chiral symmetry of the 
operators in Eqs. \eqref{eq:2.2} and \eqref{eq:2.9} 
dictate how to incorporate them in the chiral Lagrangian, 
and determine the relative importance of 
$\slashT$ couplings \cite{Mereghetti:2010tp,Maekawa:2011vs,deVries:2012ab}.
The average quark-mass term in Eq. \eqref{eq:2.2} breaks chiral symmetry as
the fourth component of an $SO(4)$ vector,
and induces interactions
proportional to powers of $\bar m r(\bar\theta)$.
The quark-mass splitting and 
the $\bar\theta$ term 
transform as different components of a single other $SO(4)$ vector,
which implies 
that their hadronic matrix elements are directly related 
\cite{Crewther:1979pi,Mereghetti:2010tp}.
They generate interactions proportional to powers of 
$\bar m \varepsilon r^{-1}(\bar\theta)$ and 
$\bar m r^{-1}(\bar\theta) (1-\varepsilon^2)\sin \bar\theta/2$, respectively.
For momenta $Q\sim m_\pi$, and taking
$\varepsilon r^{-2}(\bar\theta)={\cal O}(1)$,
quark-mass terms are paired with
chiral-symmetric operators by making $d$ count powers of the pion mass
as well. The lowest terms stemming from  Eq. \eqref{eq:2.2} are
\begin{equation}
\mathcal L^{(0,1)} =  -\frac{m^2_{\pi}}{2 D} \boldpi^2 
+\Delta m_N \left(1-\frac{2\boldpi^2}{F_\pi^2D}\right)\bar N N 
+\frac{\delta m_N}{2}
\bar N \left[\tau_3-\frac{2\boldtau\cdot\boldpi}{F_\pi D}
\left(\frac{\pi_3}{F_\pi}
+ \frac{1-\varepsilon^2}{2\varepsilon}\sin \bar\theta\right)
\right]N.
\label{eq:3.1prime}
\end{equation}
The pion mass is $m_\pi^2={\cal O}(\MQCD\bar m r(\bar\theta))$,
the nucleon sigma term is $\Delta m_N={\cal O}(m_\pi^2/\MQCD)$,
the neutron-proton mass splitting from the quark-mass difference is
$\delta m_N={\cal O}(\varepsilon r^{-2}(\bar\theta) m_\pi^2/\MQCD)$,
and pion-nucleon interactions
---$PT$ charge-symmetry breaking (CSB) for an even number of pions,
$\slashP\slashT$ for an odd number---
are determined by the matrix element that enters $\delta m_N$.
Similar relations exist at higher orders \cite{Mereghetti:2010tp}
---for example, involving the hadronic contribution to the pion mass splitting
$\delta m_{\pi}^2= {\cal O}(\delta m_N^{2})$---
but the link between $PT$ CSB and $\bar\theta$ operators
quickly becomes superfluous.

The $d=6$ operators in Eq. \eqref{eq:2.9} have different 
transformation properties under chiral symmetry,
and the connections with $PT$ operators are more tenuous.  
The qCEDM and qEDM break chiral symmetry as $SO(4)$ vectors
and induce both isospin-conserving and isospin-breaking $\slashP\slashT$ 
interactions, which are in general of the same size
because
after aligning the $\bar\theta$ term there is no longer freedom to 
eliminate the qCEDM and qEDM 
isovector components.
The important difference between qCEDM and qEDM  
is the suppression for the latter of purely hadronic operators
by powers of $\alpha_{\textrm{em}}/4\pi$.
The LR4QOs also break chiral symmetry (and isospin in particular),
but as $SO(4)$ tensors,
and the relative importance of their interactions 
is similar, but not identical, to the isovector qCEDM.
The gCEDM and PS4QOs
do not break chiral symmetry
and cannot be distinguished purely on the basis of their symmetry properties, 
more information about their matrix elements
being required.
For chiral-invariant operators, chiral-breaking interactions, 
like non-derivative pion-nucleon couplings, are suppressed by
factors of the quark masses. 

The $PT$ pion, pion-nucleon and multinucleon
chiral Lagrangians are known up to 
$\Delta = 4,3,3$, respectively 
\cite{Bijnens:1999sh,Fettes:2000gb,Ordonez:1995rz,vanKolck:1994yi}.
The 
Lagrangian from $\bar\theta$ and $d=6$
operators 
was built in great detail in Refs. 
\cite{Cheng:1990pi,Pich:1991fq,Cho:1992rv,Borasoy:2000pq,Mereghetti:2010tp,Cheng:2010rs,Maekawa:2011vs,deVries:2012ab,Bsaisou:2014oka}.
For 
nucleon and light-nuclear EDMs
at LO it is sufficient to consider a subset of the 
interactions discussed 
in Refs. \cite{Mereghetti:2010tp,Maekawa:2011vs,deVries:2012ab}: 
\begin{eqnarray}
\mathcal L_{\slashTsub} &=&  
- \frac{1}{F_{\pi}} \bar N \left(\bar g_0 \boldtau \cdot \boldpi
  + \bar g_1 \pi_3 \right) N 
- 2  \bar N \left( \bar{d}_0 
+ \bar{d}_1 \tau_3 \right) S^{\mu} N v^{\nu} F_{\mu \nu} 
\nonumber \\
& & -  \frac{\bar{\Delta}}{F_{\pi}}  \pi_3 \boldpi^2 
+  \bar C_1 \bar N N  \partial_{\mu} \left( \bar N S^{\mu} N\right) 
+ \bar C_2 \bar N \boldtau N \cdot \partial_{\mu} 
\left( \bar N S^{\mu} \boldtau N\right).
\label{eq:3.2}
\end{eqnarray}
The operators that couple a neutral pion to the vacuum (pion tadpoles)
can be eliminated order by order 
in favor of the interactions remaining in Eq. \eqref{eq:3.2}.
Each of the operators in this equation
has chiral partners, which we do not display explicitly.
The interactions in the second line are only needed at LO 
for LR4QOs, PS4QOs, and gCEDM, while 
for qEDM only EDM-type operators in the first line are important.
The coupling constants $\bar g_{0,1}$, $\bar{d}_{0,1}$, $\bar{C}_{1,2}$, and 
$\bar{\Delta}$ are discussed in the next section.

\subsection{Sizes of $\slashT$ Couplings}
\label{Sec3.2}

The LECs in the $\slashT$ hadronic Lagrangian, Eq. \eqref{eq:3.2},
are (approximately) linear functions of $\bar\theta$ and of 
the coefficients of 
$d=6$ operators in Eq. \eqref{eq:2.9}. 
We impose the constraints of chiral symmetry 
and use NDA
for an 
estimate of the scaling of the LECs.
In order to connect high-energy observables to EDMs, it is crucial that 
the coefficients of proportionality be known accurately. 
Going beyond NDA requires nonperturbative techniques, and
we review here information that can be extracted using symmetry, 
LQCD simulations, and QCD sum rules.
(A more comprehensive review of calculations
of these matrix elements can be found in, for example, 
Ref. \cite{Engel:2013lsa}.)

The LECs
$\bar g_0$ and $\bar g_1$ are isoscalar and isovector $\slashT$ 
pion-nucleon couplings, which
induce important long-range contributions to the nucleon EDM
(Section \ref{Sec4}) and to 
the $\slashT$ nucleon-nucleon potential (Section \ref{Sec5.1}).
One expects \cite{Mereghetti:2010tp,deVries:2012ab}
\begin{eqnarray}
\bar{g}_0 &=& 
{\mathcal O} \left( 
\left( \frac{m^2_{\pi}}{\MQCD^2}\bar\theta, 
\, m^2_{\pi} \tilde{d}_0, 
\, \varepsilon m^2_{\pi} \tilde{d}_3, 
\, m^2_{\pi} d_W, 
\, m^2_{\pi} \mathrm{Im}\Sigma_{a},
\, \varepsilon  \MQCD^2 \mathrm{Im}\Xi_{a}\right)   
\MQCD\right), 
\label{eq:3.3primeprime}
\\ 
\bar{g}_1 &=& {\mathcal O} \left(  
\left( \frac{\varepsilon m^4_{\pi}}{\MQCD^4}\bar\theta,
\, \frac{\varepsilon m^4_{\pi}}{\MQCD^2} \tilde{d}_0, \, m^2_{\pi} \tilde{d}_3,
\, \varepsilon m^2_{\pi} d_W, 
\, \varepsilon m^2_{\pi} \mathrm{Im}\Sigma_{a},
\, \MQCD^2 \mathrm{Im}\Xi_{a}\right)
\MQCD\right).
\label{eq:3.3prime}
\end{eqnarray}
Because $\bar{g}_{0,1}$ are LECs of chiral-breaking interactions,
for most sources the quark masses appear.
The contribution of isoscalar operators to $\bar g_1$ is suppressed by factors
of $m^2_{\pi}/M^2_{\textrm{QCD}}$, but 
as we discuss below $\bar g_1$ can still be important 
in systems, like the deuteron, 
for which the contribution of $\bar g_0$ vanishes.
Isovector operators generate both $\bar g_0$ and $\bar g_1$ at the same order, 
even though $\bar g_0$ is 
affected by the quark mass difference. 
For all the operators in Eq. \eqref{eq:2.9}, 
a third non-derivative pion-nucleon coupling, 
$(\bar g_2/F_\pi) \bar N \pi_3 \tau_3 N$, is 
suppressed, and contributes at the same level as derivative $\slashT$ couplings
\cite{Mereghetti:2010tp,deVries:2012ab}, which we neglect.  

Important consequences of chiral symmetry and vacuum alignment
are the survival of the three-pion interaction $\bar{\Delta}$ and
modifications of $\bar g_{0,1}$
proportional to it.
Both $\bar g_{0,1}$ receive tree-level contributions. Moreover, at one loop 
$\bar{\Delta}$ gives a significant contribution to the pion-nucleon FF.
Although formally an NLO effect, the loop is enhanced over NDA by a factor of
$5\pi$ \cite{deVries:2012ab}. It endows the FF with a certain
momentum dependence, whose effects on nucleon EDMs have not been studied.
For light nuclei, they were found to be small \cite{Bsaisou:2014zwa}. 
We can capture the momentum-independent effects
by redefining $\bar g_1$. Thus,
\begin{equation}
\bar g_0=  g_0
+ \frac{\delta m_N}{m^2_{\pi}} \bar \Delta
+\ldots ,
\qquad
\bar g_1=  g_1
+ 2 \left( \frac{\Delta m_N}{m^2_{\pi}}
-\frac{15 g_A^2 m_\pi}{16\pi F_\pi^2}\right)
\bar \Delta 
+\ldots ,
\label{eq:3.6new}
\end{equation}
where $g_{0,1}$ are the couplings before alignment.
The main remaining, explicit $\bar\Delta$ contribution is a tree-level
three-nucleon potential (Section \ref{Sec5.1}).
Only for LR4QOs is $\bar{\Delta}$ an LO effect, while $g_0$ is higher order.
In this case, we can eliminate
\cite{deVries:2012ab},
\begin{equation}
\bar{\Delta} = \frac{m_\pi^2}{\delta m_N}{\bar g}_0
= \mathcal O\left( \mathrm{Im}\Xi_{a} \, \MQCD^4\right).
\label{eq:3.7}
\end{equation}

The pion-nucleon couplings have been best studied 
when links to $PT$ quantities through chiral symmetry are useful:

\begin{itemize}

\item For $\bar\theta$, a comparison between Eqs. \eqref{eq:3.1prime}
and \eqref{eq:3.2}, and similarly for higher-order terms, yields
\cite{Crewther:1979pi,Mereghetti:2010tp},
\begin{eqnarray}
\frac{\bar g_0}{F_{\pi}} &=& 
\frac{\delta m_N}{F_\pi} \frac{1- \varepsilon^2}{2\varepsilon} \sin\bar\theta
=(15 \pm 2) \cdot 10^{-3} \sin\bar\theta,
\label{eq:3.6zero}
\\
\frac{\bar g_1}{F_{\pi}} - \frac{g_1}{F_{\pi}} &=&  
\left(\frac{\Delta m_N}{m^2_{\pi}} 
-\frac{15 g_A^2 m_\pi}{16\pi F_\pi^2}\right)
\frac{\delta^{} m^2_{\pi}}{F_{\pi}} 
\frac{1 -\varepsilon^2}{\varepsilon}\sin\bar\theta  
= 
- \left( 4 \pm 3 \right) \cdot 10^{-3} \sin\bar\theta,
\label{eq:3.6}
\end{eqnarray}
using the values from LQCD, 
$\varepsilon = 0.37 \pm 0.03$ \cite{Aoki:2013ldr},
$\delta m_N = 2.39 \pm 0.21$ MeV \cite{Walker-Loud:2014iea,Borsanyi:2014jba},
and $\Delta m_N = -63 \pm 9$ MeV \cite{Bali:2012qs},
and from $\chi$PT fitted to meson data,
$\delta m_{\pi}^2 = 87 \pm 55$ MeV$^2$ \cite{Amoros:2001cp}.
This value for $\bar g_0$ is somewhat smaller than the NDA estimate.
The relation between $\bar g_0$ and $\delta m_N$ is violated 
by terms of 
${\cal O}(m^2_{\pi}/M_{\textrm{QCD}}^2)$,
an effect of the same size as the uncertainty in Eq. \eqref{eq:3.6zero}. 
$g_1$ is related to 
$PT$ operators 
that contribute not to baryon masses, but to pion-nucleon scattering 
observables ---however, only at an order beyond the current 
most precise analysis \cite{Hoferichter:2009gn}.

\item 
For qCEDM, analogously \cite{Pospelov:2001ys,deVries:2012ab},
\begin{equation}
\bar g_0 = 
\delta m_N  
\left(\frac{{\tilde \delta} m_N}{\delta m_N}\frac{\tilde d_0}{\tilde c_3} 
- \frac{{\tilde \Delta}m^2_{\pi}}{m^2_{\pi}}\frac{\tilde d_3}{\tilde c_0} \right), 
\quad
\bar g_1 =  2 \, \Delta m_N  \left(\frac{{\tilde  \Delta} m_N}{\Delta m_N}
- \frac{ {\tilde \Delta} m^2_{\pi} }{m^2_{\pi}} \right) 
\frac{\tilde d_3}{\tilde c_0},
\label{g01chromo}
\end{equation}
where $\tilde c_{0,3}$ are the coefficients of chromomagnetic operators 
analogous to the chromoelectric operators in Eq. \eqref{eq:2.9},
${\tilde \Delta} m^2_{\pi}$, ${\tilde \Delta} m_N$, 
and ${\tilde \delta} m_N$ are corrections to
the pion mass, nucleon sigma term, and nucleon mass splitting due to 
$\tilde c_{0}$, $\tilde c_{0}$, and $\tilde c_{3}$, respectively. 
The evaluation of the corresponding matrix elements is 
currently being pursued by lattice collaborations 
\cite{WalkerLoud:2015}.
At the moment, the best estimates on $\bar g_{0,1}$ come from QCD sum rules. 
It is found \cite{Pospelov:2005pr} that $\bar g_1$ 
is larger than $\bar g_0$ by a factor of about 5, and 
\begin{equation}
\frac{\bar g_1}{F_{\pi}} = 
- ( 20 ^{+40}_{-11} )  \cdot 10^{-3} \; (2\pi F_\pi)^2 \tilde{d}_3 .
\label{g1qCEDMSR}
\end{equation}
The large error is due to cancellations 
in Eq. \eqref{g01chromo}, 
which complicate estimates of these couplings. 
In terms of dimensionless quantities, the numerical factors in 
Eqs. \eqref{eq:3.6zero} and \eqref{g1qCEDMSR} are not very different, 
as one would expect from Eqs. \eqref{eq:3.3primeprime} and \eqref{eq:3.3prime}.

\end{itemize}

The parameters $\bar d_{0,1}$ represent 
short-range contributions to the nucleon EDM
(Section \ref{Sec4}) 
and have expected sizes \cite{Mereghetti:2010tp,deVries:2012ab}
\begin{equation}
\bar{d}_{0,1} =
{\mathcal O}\left( 
\left( \frac{m^2_{\pi}}{\MQCD^2}\bar\theta, 
\, m^2_{\pi}  \tilde{d}_i,
\, m^2_{\pi}  d_i,
\, \MQCD^2  d_W, 
\, \MQCD^2  \mathrm{Im}\Sigma_{a},
\, \MQCD^2  \mathrm{Im}\Xi_{a}
\right)  
\frac{e}{\MQCD}
\right).
\label{eq:3.3primeprimeprime}
\end{equation}
Chiral breaking intrinsic to the electromagnetic interaction ensures
that no extra factors of quark masses are needed beyond those appearing
explicitly in Eqs. \eqref{eq:2.2} and \eqref{eq:2.9}. 
Because some of this breaking involves isospin,
all sources induce isoscalar and isovector
components of the same size. 
$\bar d_{0,1}$ are not fixed by symmetry but
can be extracted from LQCD calculations of the
nucleon EDMs, as we discuss in Sect. \ref{Sec4.2}.

The remaining couplings in 
Eq. \eqref{eq:3.2}, $\bar C_{1,2}$, 
are less well known, and have not been studied extensively.
They represent short-distance contributions to the nucleon-nucleon 
potential and their main phenomenological impact is through light-nuclear EDMs 
(Section \ref{Sec5}).
They are chiral-invariant interactions that appear 
without $m_\pi^2/\MQCD^2$ suppression,
and at LO, only for chiral-invariant operators
\cite{deVries:2012ab},
\begin{eqnarray}
\bar C_{1,2}  =  \mathcal O\left( 
\left(d_W,
\mathrm{Im}\Sigma_{a}\right) 
\frac{\MQCD}{F^2_{\pi}}  \right).
\label{eq:3.6prime}
\end{eqnarray}
For this reason, contact terms with different isospin structures are not 
needed at LO.

Equations \eqref{eq:3.3primeprime}--\eqref{eq:3.6prime}
show how the different properties of $\bar\theta$ and $d=6$
operators under chiral symmetry 
lead to 
different hierarchies among the couplings in 
Eq. \eqref{eq:3.2}, which has profound consequences for the EDM of the 
nucleon and light nuclei.
Taking into account Eq. \eqref{eq:3.7},
these observables are expected to be described in LO in terms
of the six independent LECs $\bar g_{0,1}$, $\bar d_{0,1}$, 
and $\bar C_{1,2}$.
In Section \ref{Sec4} we find  that
the nucleon EDM depends mostly on $\bar g_{0,1}$ and $\bar d_{0,1}$,
whereas
in Section \ref{Sec5} we show 
that the six LECs enter nuclear EDMs. 
By contrast, the traditional
description of nuclear $\slashT$ is largely based on 
the three non-derivative pion couplings \cite{Barton:1961eg},
$\bar g_{0,1}$ and $\bar g_{2}$, which is subleading for all
$d=6$ sources.

\section{Peccei-Quinn Mechanism}
\label{Sec2.3}

So far, we have assumed that there is no particular reason behind the 
small value of $\bar\theta$.
A very elegant way to obtain a small $\bar\theta$ dynamically
is the PQ mechanism \cite{Peccei:1977hh},
which is discussed in EFT in Ref. \cite{Georgi:1986df}.
The approximate, spontaneously broken $U(1)_{PQ}$ symmetry 
is realized nonlinearly, with SM fields invariant and the pseudo-Goldstone
boson, the axion $a$ \cite{Weinberg:1977ma,Wilczek:1977pj}, changing
by an additive constant, $a \rightarrow a + c$. 
The symmetry is explicitly broken by the anomalous coupling to 
$G \tilde G$, which can be eliminated,
as we did in Section \ref{Sec2.1}, in favor of a complex quark mass term
---except that now the axial rotation depends on $a$.

After vacuum alignment, which in this context is equivalent to the 
diagonalization of the pion-axion mass term,
the axion Lagrangian reads
\begin{eqnarray}
\mathcal L_{ax} & = &  \frac{1}{2} \partial_{\mu} a \partial^{\mu} a 
+ \frac{1}{2f_a}\bar q \left[C_0
+ \left(C_1+\varepsilon \, r^{-2}\! 
\left(\bar\theta + a/f_a\right)\right)\tau_3\right]
\gamma_5 \gamma_{\mu} q\, \partial^{\mu} a 
\label{eq:2.19}\\
&&
- \bar m \, r\! \left(\bar\theta + a/f_a\right) \bar q q 
+ \varepsilon \bar m \, r^{-1}\!\left(\bar\theta + a/f_a\right)
\bar q \left[\tau_3 +\frac{(1-\varepsilon^2)}{2\varepsilon} 
\sin\! \left(\bar\theta + a/f_a\right) i \gamma_5  \right]  q 
+\ldots
\nonumber 
\end{eqnarray}
where  $f_a$ is the axion decay constant and 
the two couplings $C_{0,1}$ are model dependent.
When chiral symmetry is broken, the term proportional to $\bar q  q$ generates 
the axion potential $V(\bar\theta + a/f_a)$, 
whose minimum (curvature) determines the axion vev $\langle a \rangle$
(mass). If $\bar\theta$ is the only source of $\slashCP$, the potential is minimized by $\bar\theta + \langle a \rangle/f_a = 0$.
The presence (in the ``$\dots$'') of higher-dimensional $\slashCP$ operators
that break chiral symmetry affects this potential; for example for
the qCEDM,
\begin{equation}
V(x) = 
- \frac{m_{\pi}^2 F_{\pi}^2}{4} \, r(x) 
\left[1- \frac{\tilde{\Delta}m^2_{\pi}}{2\tilde c_0m_{\pi}^2 } 
\left(\tilde d_0 + \varepsilon \tilde d_3 \right) \, r^{-2}(x)\, \sin x\right].
\label{eq:2.22}
\end{equation}
Minimization results in an induced angle
\begin{equation}
\bar\theta_\textrm{ind} =  \bar\theta + \frac{\langle a \rangle}{f_a}
= \frac{2}{1-\varepsilon^2} \frac{\tilde{\Delta}m^2_{\pi}}{\tilde c_0 m^2_{\pi}}   
\left( \tilde d_0 + \varepsilon \, \tilde d_3   \right)
\label{eq:2.23}
\end{equation}
of ${\cal O}(M_{\textrm{QCD}}^2/M^2_{\slashTsub})$, similar 
to other 
$d=6$ operators. 
The consequence to low-energy dynamics is that the coupling $\bar g_0$ induced 
by the qCEDM receives another correction, 
\begin{equation}
(\bar g_0)_{\textrm{PQ}} 
= \bar g_0 + \delta m_N \frac{1-\varepsilon}{2\varepsilon}
\,\bar\theta_{\textrm{ind}}
= \frac{\delta m_N}{\varepsilon \tilde c_3} 
\left(\varepsilon   +  \frac{\tilde c_3}{\tilde c_0 }  
\frac{ \tilde{\Delta}  m^2_{\pi} }{m^2_{\pi}}  \right) \tilde d_0,
\end{equation}
which cancels the contribution from $\tilde d_3$.
Similar relations can be worked out for LR4QOs, resulting
in a vanishing $\bar g_0$.

\section{Nucleon Electric Dipole Moment}
\label{Sec4}

The study of the nucleon EDM in $\chi$PT has a long history
starting with 
Ref. \cite{Crewther:1979pi},
where the leading pion-loop contribution to the neutron EDM 
induced by 
$\bar\theta$ was computed.
The calculation was later extended to the 
radius \cite{Thomas:1994wi} 
and then full \cite{Hockings:2005cn} electric dipole form factor (EDFF), 
and to 
NLO \cite{Narison:2008jp,Ottnad:2009jw,Mereghetti:2010kp}. 
The nucleon EDFF generated by the $d=6$
operators 
was computed to NLO 
in Refs. \cite{deVries:2010ah,deVries:2012ab}.
In addition to the EDM, the momentum dependence of the EDFF is interesting. 
Due to Schiff's theorem \cite{Schiff:1963zz}, 
the nucleon EDM does not contribute to the EDM 
of atoms in the nonrelativistic limit, and the first non-vanishing contribution is induced by the 
EDFF radius.
Furthermore, 
the EDFF presented in Section \ref{Sec4.1} can be used
to guide the extrapolation of LQCD results in both
pion mass \cite{O'Connell:2005un,Guo:2012vf,Akan:2014yha}
and 
momentum,
as described in Section \ref{Sec4.2}.
The 
role of strangeness is examined in Section \ref{Sec3.3}.

\subsection{Chiral EFT}
\label{Sec4.1}

$\chi$PT allows for the calculation of low-energy observables in a
controlled perturbative expansion in powers of $Q/\MQCD$, where
each loop contributes $Q^2/\MQCD^2$ \cite{Weinberg:1978kz}.
A review,
including the $PT$ electric charge and magnetic dipole FFs, can be found in
Ref. \cite{Bernard:1995dp}.
The $\slashP T$ toroidal dipole FF can be found in 
Refs. \cite{Maekawa:2000qz,Maekawa:2000bd}.
The $\slashP \slashT$ component of the electromagnetic current can be written as
\cite{Hockings:2005cn,deVries:2010ah}
\begin{equation}
J^\mu_{\slashTsub}(q,K)= 2  \left(F_0(Q^2)+F_1(Q^2) \tau_3\right) 
\left[S^\mu v\cdot q- S\cdot q v^\mu
+\frac{1}{m_N}\left(S^{\mu} q\cdot K -S\cdot q K^{\mu}\right)+\ldots\right],
\label{eq:4.1}
\end{equation}
where $q=p- p'$ and $K=(p+p')/2$ in terms of the nucleon momentum in
the initial (final) state $p$ ($p'$), and $Q^2 = -q\,^2>0$.
Here $F_0(Q^2)$ ($F_1(Q^2)$) is the isoscalar (isovector) EDFF of the nucleon, 
\begin{equation}
F_i(Q^2)=d_i - S'_i \; Q^2 + H_i(Q^2), 
\label{eq:4.2}
\end{equation}
where  $d_i$ is the EDM, $S'_i$ the Schiff moment, and $H_i(Q^2)$  
accounts for the remaining  $Q^2$ dependence.  

\begin{figure}
\center
\includegraphics[width = 12cm]{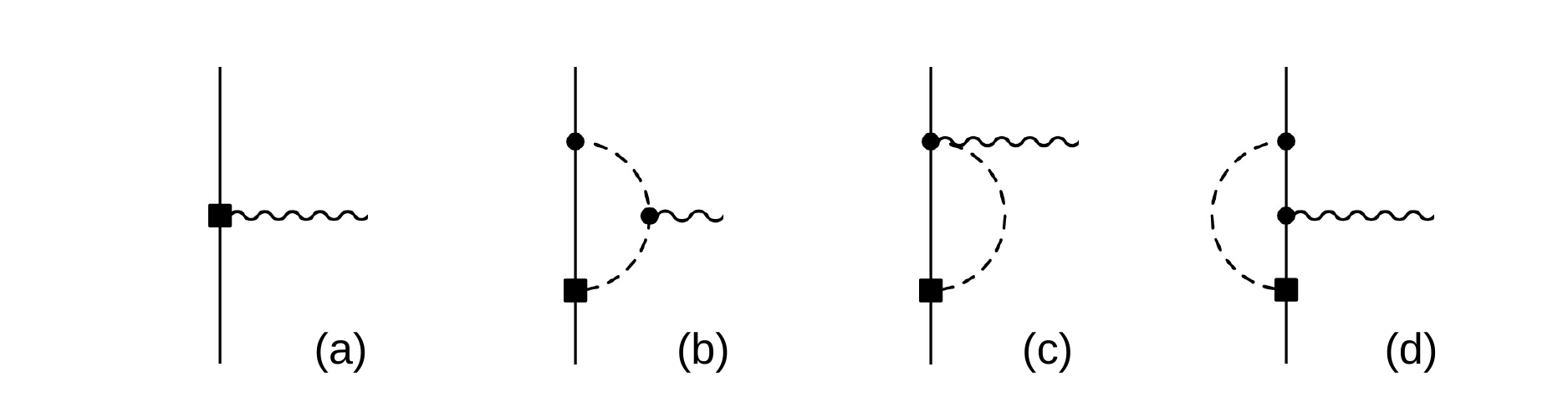}
\caption{Leading diagrams for the nucleon EDM. 
Solid, dashed and wavy lines represent propagation of
nucleons, pions and photons, 
respectively.
Dots (squares) denote $T$ ($\slashT$) interactions. 
For simplicity, only one possible ordering is shown.}
\label{NEDMLO}
\end{figure}

For all $\slashT$ sources, to NLO
\cite{Crewther:1979pi,Narison:2008jp,Ottnad:2009jw,deVries:2010ah,Mereghetti:2010kp,deVries:2012ab}
\begin{eqnarray}
\frac{d_n+d_p}{2} &=& \bar{d}_0
        +\frac{eg_A\bar{g}_0}{(2\pi F_{\pi})^2}\; \pi
        \left[\frac{3m_{\pi}}{4m_N}
        \left(1+\frac{\bar{g}_1}{3\bar{g}_0}\right) - \frac{\delta m_N}{m_{\pi}}
        \right],
\label{eq:4.3}\\
\frac{d_p-d_n}{2} &=& \bar{d}_1 
        +\frac{eg_{A}\bar{g}_{0}}{(2\pi F_{\pi})^2}
        \left[L-\ln\frac{m_{\pi}^2}{\mu^2}
        +\frac{5\pi}{4}\frac{m_\pi}{m_N}
         \left(1+\frac{\bar{g}_1}{5\bar{g}_0}\right) 
        - \frac{\hat{\delta} m^2_{\pi}}{m^2_{\pi}}\right],
\label{eq:4.4}
\end{eqnarray}
where $L = 2/(4-d) - \gamma_E + \log 4\pi$, with $d$ the spacetime dimension
and $\gamma_E=0.557...$ the Euler constant.
For illustration, the leading diagrams are shown  
in Fig. \ref{NEDMLO}.  
The nucleon magnetic dipole moment (MDM) couples to the electric field
in the $\Delta=2$ Lagrangian, an effect $\propto m_N^{-2}$.
It does not contribute to the EDM before next-to-next-to-leading order
(N$^2$LO), contrary to 
uncontrolled calculations based on a ``relativistic'' chiral
Lagrangian \cite{He:1989xj}. 
The subset of $\mathcal O(1/m_N^{2})$ corrections that are proportional to 
the 
MDM is small \cite{Seng:2014pba}.

{}From the NDA estimates of Section \ref{Sec3.2}, we see that 
the relative importance of the terms in Eqs. \eqref{eq:4.3} 
and \eqref{eq:4.4} depends on the source:
\begin {itemize}

\item
${\bar d}_{0,1}$, Fig. \ref{NEDMLO}(a), dominates, while
loop contributions, Fig. \ref{NEDMLO}(b-d), appear at least at N$^2$LO
for qEDM, gCEDM, and PS4QOs. 
The full set of N$^2$LO corrections 
was considered in Ref.  \cite{deVries:2010ah}.

\item
$\bar g_0$ appears at LO for sources that break 
chiral symmetry and do not contain photon fields: 
$\bar\theta$, qCEDM and LR4QOs. 
Its one-loop contribution is purely isovector, 
and the dependence on $L$ and $\mu$ is eliminated solely by ${\bar d}_1(\mu)$.
Non-analytic corrections to the isoscalar EDM are finite and suppressed by 
$m_{\pi}/m_N$. Despite the extra factor of $\pi$ with respect to NDA,
they are about 10\% of the leading loop. 
However, one cannot exclude,
on the basis of 
Eq. \eqref{eq:3.3primeprimeprime}, 
that $\bar d_0$ is present at the same order. 
Without further dynamical information, one cannot state, as sometimes
in the literature, that the nucleon EDM is isovector at this order. 

\item
$\bar g_1$ contributes
through recoil corrections in the nucleon propagator and pion-nucleon axial 
couplings, and only affects the proton EDM.
In the case of isoscalar sources, $\bar g_1$ is formally suppressed by 
$m_{\pi}^2/\MQCD^2$. 
Even for $\bar\theta$, where $\bar g_0$ is a factor of ten smaller 
than expected by power counting, 
$\bar g_1$ gives a correction to the proton EDM 
which is only a few percent of the leading loop. 
In the case of isospin-breaking sources, 
$\bar g_{0,1}$ 
appear at LO, but
$\bar g_0$ arises from pion tadpoles and 
is proportional to $\delta m_N$, which makes it numerically smaller than  
$\bar g_1$ \cite{deVries:2012ab}. 
For the isovector qCEDM and LR4QO, then, the nucleon EDM 
is likely to receive its largest numerical contribution from the 
LECs $\bar d_{0,1}$, with non-analytic terms entering at the $30\%$ level.
\end{itemize}

For concreteness, we specify Eqs. \eqref{eq:4.3} and \eqref{eq:4.4} 
for $\bar\theta$, where $\bar g_0$ is well determined  
by Eq. \eqref{eq:3.6zero} and $\bar g_1$ gives only an N$^3$LO effect.
Setting the renormalization scale $\mu = m_N$ and 
neglecting the numerically small isospin-breaking contributions,
\begin{eqnarray}
d_n -(\bar d_0 -\bar d_1)&=&
-\frac{eg_{A} \bar g_0}{(2\pi F_{\pi})^2}
\left[\ln\frac{m_N^2}{m^2_{\pi}} + \frac{\pi}{2}\frac{m_{\pi}}{m_N}\right] 
\bar\theta 
= 
- (1.99 + 0.12) \cdot 10^{-3} \sin \bar\theta \; e \, {\rm fm}, 
\label{eq:4.6}\\
d_p - (\bar d_0 +\bar d_1)&=& 
\frac{eg_{A} \bar g_0}{(2\pi F_{\pi})^2}
\left[ \ln\frac{m_N^2}{m^2_{\pi}} + 2\pi \frac{m_{\pi}}{m_N} \right] \bar\theta 
= 
(1.99 + 0.46  ) \cdot 10^{-3} \sin \bar\theta \; e \, {\rm fm} ,
\label{eq:4.7} 
\end{eqnarray}
showing that, especially for the neutron, 
convergence of the $SU(2)$ chiral expansion is good.
Assuming that there are no fine-tuned cancellations \cite{Crewther:1979pi}
between $\bar d_{0,1}$, 
which are analytic in $m_{\pi}^2$, and non-analytic contributions, 
the current bound on  
$d_n$ \cite{Baker:2006ts} 
allows to put a bound on $ \bar\theta$.
Similarly, using the NDA expressions for $\bar g_{0,1}$ and $\bar d_{0,1}$ 
given in Section \ref{Sec3.2} and assuming no cancellations, one can 
bound the coefficients of 
$d=6$ operators \cite{deVries:2010ah}:
\begin{equation}
\bar\theta, \, \MQCD^2 \tilde d_{i}, \, \MQCD^2 d_{i} \,  \simle 10^{-10}, 
\quad 
\MQCD^2 d_W, \, \MQCD^2 \mathrm{Im}\Sigma_{a}, \, \MQCD^2 \mathrm{Im}\Xi_{a} 
\simle 10^{-12}.
\label{eq:4.8}
\end{equation}
The weaker bound on $ \bar\theta$, qCEDM and qEDM reflects
the proportionality to light-quark masses.
If the dimensionless $\slashT$ parameters in Eq. \eqref{scalingofdim6}
are ${\cal O}(1)$,
the bounds in Eq. \eqref{eq:4.8} 
suggest new physics at a scale $M_{\slashTsub}\simge 100$ TeV. 
Once NDA can be replaced by more accurate determinations of matrix elements
in LQCD, these bounds 
can be reliably translated into bounds on the dimensionless parameters 
at $M_{\slashTsub}$ 
using the RG of Section \ref{Sec2.2}, as sketched in Ref. \cite{Dekens:2013zca}.

While certainly an exciting evidence of new physics, 
the observation of neutron \textit{and} proton EDMs
would not, with the current theoretical status,  
be sufficient to clearly identify the source of new physics,
and in particular to disentangle the effect of $\bar\theta$ 
from higher-dimension operators.
Without further input from nonperturbative techniques, 
Eqs. \eqref{eq:4.6} and \eqref{eq:4.7} show that,
for all operators 
in Eq. \eqref{eq:2.9}, $d_{p,n}$
contain at least two unknown LECs.
An interesting independent observable is 
the momentum dependence of the EDFF. 
To NLO 
\cite{Thomas:1994wi,Hockings:2005cn,Mereghetti:2010kp,deVries:2010ah,deVries:2012ab},
the square radii are
\begin{equation}
S^{\prime}_0 = -\frac{\pi e g_A \bar g_0\delta m_N}{12 (2\pi F_{\pi})^2 m_{\pi}^3},
\qquad
S^{\prime}_1 = \frac{e g_A \bar g_0}{6 (2\pi F_{\pi})^2 m_{\pi}^2} 
\left(1 - \frac{5 \pi}{4} \frac{m_{\pi}}{m_N}
- \frac{\hat{\delta} m_\pi^2}{m_\pi^2}\right) .
\label{NSM}
\end{equation}
For $\bar\theta$, qCEDM and LR4QO, radii arise at
the same order as the EDM, are finite in the chiral limit
and approximately isovector.
They are dominated by contributions at the scale $m_{\pi}$, which are 
known except for the $\slashT$ parameters.
For example, for $\bar\theta$ \cite{Mereghetti:2010kp},
\begin{equation}
S^{\prime}_0 = -5.0 \cdot 10^{-6} \sin \bar \theta \; e \, \mathrm{fm}^3,
\qquad
S^{\prime}_1 = 6.8 \cdot 10^{-5} \sin \bar \theta \; e \, \mathrm{fm}^3.
\label{NSM2}
\end{equation}
In contrast, for qEDM, gCEDM and PS4QOs, radii 
arise at N$^2$LO and scale as 
$Q^{2}/\MQCD^2$ with respect to the EDM \cite{deVries:2010ah}.
The functions $H_i(Q^2)$ from Eq. \eqref{eq:4.2}
can be found, to NLO, in 
Refs. \cite{Hockings:2005cn,Mereghetti:2010kp,deVries:2010ah,deVries:2012ab}.
Although an experimental measurement of the momentum dependence of the EDFF 
is not going to happen any time soon, 
the extrapolation 
$Q^2\rightarrow 0$
of LQCD EDFFs for $\bar\theta$ and qCEDM
would allow to extract 
$\bar g_0$ and $\bar d_{0,1}$ at the same time.

\subsection{Interplay with Lattice QCD}
\label{Sec4.2}

The best tool to determine the LECs $\bar d_{0,1}$ 
that
contribute to the nucleon EDM is LQCD.
Unfortunately there are virtually no LQCD results for the
$d=6$
sources, most work having focused on $\bar\theta$.
A collection of early results can be found in Ref. \cite{Lin:2011cr}.

\begin{figure}
\center
\includegraphics[width = 10cm]{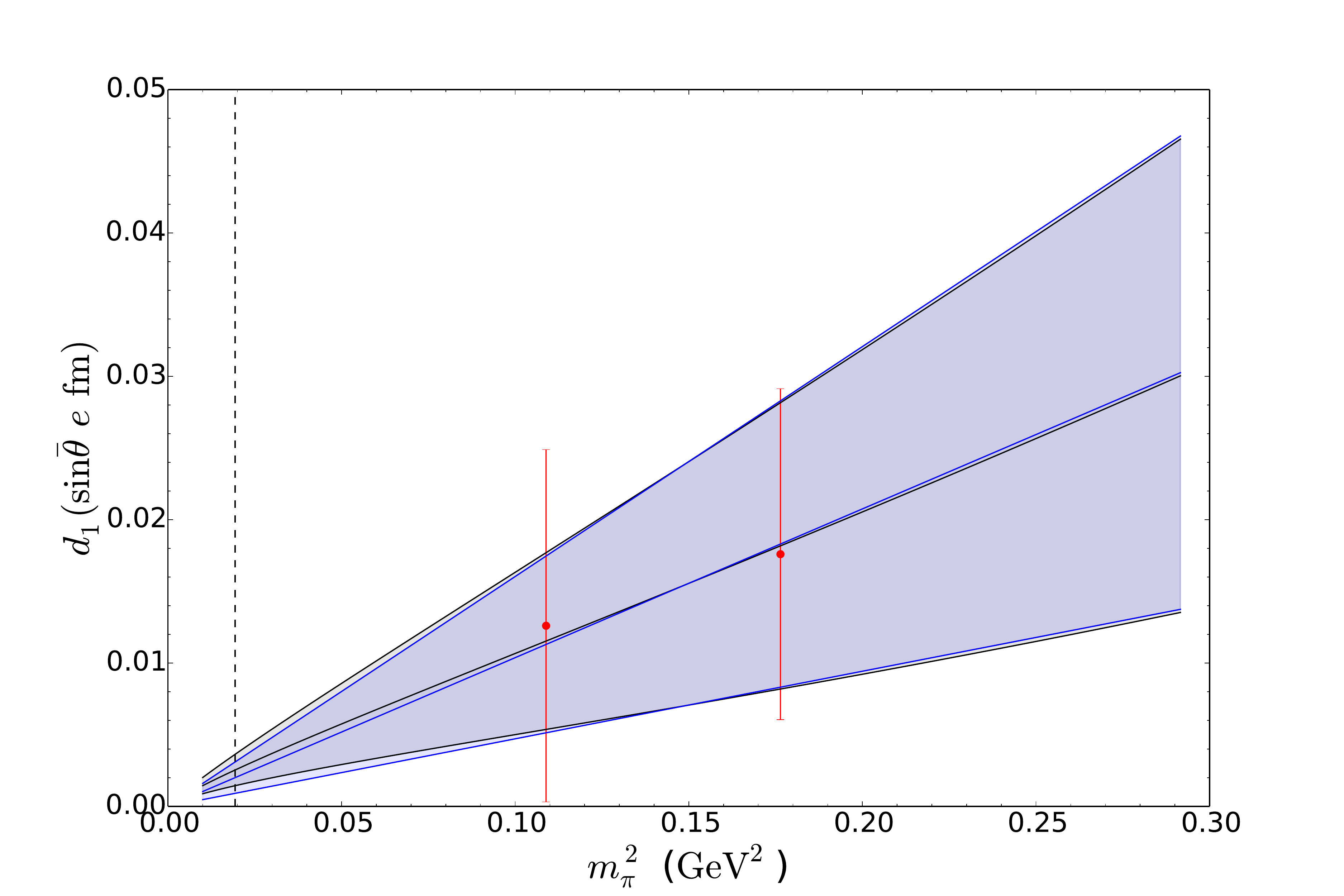}
\caption{Isovector component $d_1$ of the nucleon EDM 
(in units of $\sin{\bar\theta}\, e$ fm) as function
of the pion mass squared $m_\pi^2$ (in units of GeV$^2$).
The black lines and gray shaded area show the best fit 
and $1\sigma$ uncertainty obtained by fitting Eq. \eqref{eq:4.4}
to the 
LQCD results of Ref. \cite{Shintani:2014zra}, represented by the red
points with error bars. 
The blue lines and blue shaded area are the corresponding fit
with $\bar g_0 = 0$. The dashed vertical line indicates the physical pion mass.}
\label{Fig3}
\end{figure}

A recent LQCD evaluation of 
$d_{p,n}$ from
$\bar\theta$ can be found in Ref. \cite{Shintani:2014zra}.
Simulations were performed at two values of the pion mass, 
$m_{\pi} = 330, 420$ MeV, with domain-wall fermions to suppress 
the lattice artifact of chiral symmetry violation. 
Results, which still have a significant statistical error, 
are extrapolated linearly to $Q=0$, 
where they give 
EDMs compatible with zero.
Ref. \cite{Akan:2014yha} improved the finite-volume corrections
of Refs. \cite{O'Connell:2005un,Guo:2012vf} and used these 
LQCD results to extract 
$\bar d_{0,1}$
with $SU(3)$ $\chi$PT at NLO.
We have 
fitted the same EDM points but with the $SU(2)$ formulas, 
Eqs. \eqref{eq:4.3} and \eqref{eq:4.4}, and
neglecting finite-volume corrections.
We find good fits, as exemplified for the isovector component 
in Fig. \ref{Fig3},
where 
the best fit results
and $1\sigma$ uncertainty obtained from Eq. \eqref{eq:4.4}, 
with $\bar g_0$ fixed to the value \eqref{eq:3.6zero},
are compared 
with results of fitting $\bar d_1$ only and
neglecting the chiral loop, that is, setting $\bar g_0$ to zero. 
Clearly, lattice EDM data show an essentially linear dependence 
on $m_\pi^2$ with no clear sign of the chiral log.
For the short-range LECs at the scale $\mu = 939$ MeV, we find,
factoring out the pion-mass dependence, 
\begin{eqnarray}
\bar d_{0}(\mu ) = \left( -0.04 \pm 0.45 \right) 
\frac{m_{\pi}^2}{(2\pi F_{\pi})^3}\, e \sin \bar\theta, 
\quad  \bar d_{1}(\mu) = \left( 0.05 \pm 0.45 \right)  
\frac{m_{\pi}^2}{(2\pi F_{\pi})^3} \, e \sin \bar\theta,
\end{eqnarray}
which is not far from the  NDA estimate of Section \ref{Sec3.2}, 
once one considers the large uncertainties.
Extrapolating to the physical pion mass, 
again using Eq. \eqref{eq:3.6zero},
\begin{equation}
d_n = -( 2.4 \pm 1.5 ) \cdot 10^{-3} \sin\bar\theta \, e \, \textrm{fm}, \qquad
d_p =  ( 2.5 \pm 1.5 ) \cdot 10^{-3} \sin\bar\theta \, e \, \textrm{fm},
\label{latticenuc}
\end{equation}
which are consistent with the more sophisticated analysis of 
Ref. \cite{Akan:2014yha}.
After this review was completed,
an LQCD calculation with clover fermions and imaginary
$\bar\theta$ has appeared \cite{Guo:2015tla}, which gives a similar value 
for $d_n$ but with smaller error bars .

Eventually, the extrapolation in $Q^2$ could also be made with 
the full $\chi$PT EDFF, Eq. \eqref{eq:4.2}
\cite{Hockings:2005cn,Mereghetti:2010kp,deVries:2010ah,deVries:2012ab}.
The range in $Q$ used in  Ref. \cite{Shintani:2014zra} (from about 450 to 
800 MeV) is unfortunately beyond the validity of $\chi$PT,
but new calculations, already under way 
at a lighter pion mass ($m_{\pi} = 170$ MeV), 
are performed at smaller $Q^2$ \cite{Shintani:2015}.
They should make it possible to extract $\bar g_0$ by fitting the 
slope of the EDFF with the $\chi$PT prediction in Eq. \eqref{NSM}, 
and to check the estimate in Eq. \eqref{eq:3.6zero}.

The uncertainty on the LQCD evaluation of $\bar d_{0,1}$
is quite large and
compatible with zero. 
Nonetheless, for $\bar\theta$ LQCD is now competitive with other 
nonperturbative techniques.
For comparison, 
from QCD sum rules
\cite{Pospelov:1999mv,Pospelov:1999ha,Hisano:2012sc},
\begin{equation}
|d_n| = \left( 2.5 \pm 1.3 \right) \cdot 10^{-3} \bar\theta\, e\, \textrm{fm}.
\end{equation}
QCD sum rules also give the best available estimate 
for the qCEDM \cite{Pospelov:2000bw},
\begin{equation}
|d_n | = 
\left(1 \pm 0.5 \right) (2\pi F_\pi)^2 
| 0.9 \, \tilde d_0 
-  0.3 \, \tilde d_3 |    \cdot 10^{-3} \, e\, \textrm{fm}. 
\end{equation}

Among $d=6$ sources, the qEDM is particularly simple,
since to a good approximation $d_{p,n}$ arise solely from
${\bar d}_{0,1}$ with small isospin-breaking corrections.
{}From Eqs. \eqref{eq:2.9} and \eqref{eq:3.2}, the photon
couples in LO to the isoscalar and isovector tensor charges 
defined by
$\langle N|{\bar q}\sigma_{\mu\nu} q|N \rangle=
4g^T_{0} \epsilon^{\mu\nu\rho\sigma}v_\rho{\bar N}S_{\sigma} N$,
$\langle N|{\bar q}\sigma_{\mu\nu}\tau_3 q|N \rangle=
4g^T_{1} \epsilon^{\mu\nu\rho\sigma}v_\rho{\bar N}S_{\sigma}\tau_3 N$,
resulting in 
\begin{eqnarray}
\frac{d_n+d_p}{2} &=& 
g^T_{0} \bar m \, e  \left(\frac{d_0}{3} + d_3 \right) 
= ( 0.29 \pm 0.03 ) \, \bar m \, e \left(\frac{d_0 }{3} + d_3 \right) ,
\label{tensor2}\\
\frac{d_p-d_n}{2} &=& 
g^T_{1} \bar m \, e \left(d_0 + \frac{d_3}{3} \right) 
= ( 0.51 \pm 0.04) \, \bar m \, e \left(d_0 + \frac{d_3}{3}\right),
\label{tensor1}
\end{eqnarray}
where we used the recent LQCD results \cite{Bhattacharya:2015,Gupta:2015tpa,Bhattacharya:2015gma} 
for the proton tensor charges at $\mu = 2$ GeV,
and quark mass and 
qEDM should also be evaluated at this scale.
These numbers are in good agreement with the NDA expectation.
Existing extractions of the proton tensor charges from experiment
\cite{Bacchetta:2012ty,Anselmino:2013vqa} are at too high a $Q^2$ for $\chi$PT.
Some of the issues affecting the lattice 
implementation of higher-dimensional operators are discussed in Ref. 
\cite{Bhattacharya:2015rsa}.

Our simplified analysis of LQCD data, which omitted correlations,
finite-volume effects and systematic errors, suggests that there can be 
a rich interplay between LQCD and $\chi$EFT.
It will be interesting to see whether the 
substantial reduction 
of the LQCD uncertainty to the 10\% level expected in the near future 
for $\bar\theta$ \cite{Shintani:2014zra}
will show a clear signal of an EDM and the presence of a chiral logarithm,
allowing an extraction of $\bar g_0$.
More generally, LQCD calculations of matrix elements for other
sources would fill a gaping hole in the road from BSM physics
to EDM experiments.

\subsection{Role of Strangeness}
\label{Sec3.3}

Much of the low-energy $\slashT$ literature includes the strange quark $s$
explicitly. Yet, 
the intermediate value of 
its mass $m_s$ poses a problem:
it prevents integrating $s$ out at a perturbative scale, as it is done for 
the $t$, $b$ and, to a lesser extent, $c$ quarks,
but leads to poor convergence of the $\chi$PT expansion
via a relatively large kaon mass $m_K$ \cite{Donoghue:1998bs}.
Although the best way to deal with 
$s$ (and maybe also $c$) is through 
nonperturbative techniques,
symmetry considerations also
give some insight.

$\bar\theta$ effects
in $SU(3)_L \times SU(3)_R$
($U(3)_L \times U(3)_R$ if one wants an explicit connection to the 
$U(1)_A$ anomaly) $\chi$PT 
\cite{Cheng:1990pi,Pich:1991fq,Cho:1992rv,Borasoy:2000pq,Cheng:2010rs}
are proportional to
\begin{equation}
\frac{m_d m_u m_s}{m_s (m_d + m_u) + m_u m_d} = 
\frac{\bar m}{2}  (1-\varepsilon^2)
\left[ 1 + \frac{\bar m}{m_s} (1-\varepsilon^2) + \ldots \right]. 
\end{equation}
Although $\bar g_0$ and $\delta m_N$ receive, starting at order 
$\mathcal O( m_K/M_{\textrm{QCD}})$, large corrections 
from kaon and eta meson loops (which show little sign of convergence),
these corrections affect $\bar g_0$ and $\delta m_N$ in exactly the same way
up to N$^2$LO and are already accounted for through
Eq. \eqref{eq:3.6zero} when extracting $\delta m_N$ 
from LQCD simulations with dynamical strange quarks. 
At N$^2$LO, in both $SU(2)$ and $SU(3)$ $\chi$PT there are short-distance 
contributions to $\bar g_0$ that do not affect $\delta m_N$,
but, since they are not proportional to $m_s$,
their numerical effect should not be larger than  
the uncertainty quoted in Eq. \eqref{eq:3.6zero}.
The isospin-violating coupling $\bar g_1$ receives a tree-level
contribution proportional to the $\pi$-$\eta$ mixing angle,
which, despite being formally LO, 
is suppressed with respect to $\bar g_0$ by $\bar m/m_s$ and is numerically
of the same size as the estimate of Eq. \eqref{eq:3.6}. 
The first contribution 
not suppressed by powers of $\bar m/m_s$ appears at N$^2$LO, 
as in $SU(2)$ $\chi$PT. 
Thus, strangeness should not significantly affect the values 
of the pion-nucleon couplings from $\bar\theta$.

The nucleon EDM has been computed in this framework to various degrees
of sophistication 
\cite{Crewther:1979pi,Ottnad:2009jw,Guo:2012vf}.
At one loop, there are 
additional nucleon-kaon and nucleon-eta couplings,
which at LO are fixed by combinations of baryon masses, for example  
$\bar g_{N \Sigma K}$ is related to the mass difference between the nucleon and 
the $\Sigma$ baryon, $m_{N} - m_{\Sigma}$. 
As for $\bar g_0$, these relations are violated at N$^2$LO. However, 
now the corrections are proportional to $m_s$, and are numerically more 
important \cite{deVries:2015}. 
For the nucleon, pion and kaon loops give contributions of approximately 
the same size. However, NLO corrections to the kaon contributions are as 
large as LO, casting some doubts on the convergence of the $SU(3)$ expansion.
On the contrary, the $SU(2)$ expansion of the nucleon EDM converges fairly well.

In the 
$d=6$ sector, the basis in Eq. \eqref{eq:2.9} 
needs to be enlarged to include $s$.
Four additional quark bilinears can be constructed, 
the strange EDM (sEDM) and CEDM (sCEDM), and two $\Delta S = 1$ flavor-changing neutral 
currents of the same form as the qEDM and qCEDM but containing a $d$ 
and an $s$. 
There are also several more four-quark operators, listed 
in Ref. \cite{An:2009zh}, which, however, does not fully 
account for the constraint of SM gauge symmetry.
If one considers only $\Delta S = 0$ operators, 
in addition to the four operators already defined in Eq. \eqref{eq:2.9},
eight four-quark operators with two strange quark fields receive 
non-vanishing matching coefficients at tree level, so that only half 
of the operators defined  Ref. \cite{An:2009zh} needs to be considered.
The matrix elements of these operators were studied \cite{An:2009zh}
in the factorization approximation and in the quark model, 
but more rigorous techniques are needed.

The sEDM and sCEDM have received more attention.
The effect of the sEDM $d_s$ on 
$d_{p,n}$ can be obtained from
$ \langle N | \bar s \sigma^{\mu \nu} s | N \rangle  
= 2 g^T_{s} \epsilon^{\mu\nu\rho\sigma}v_\rho{\bar N}S_{\sigma} N$ on the lattice:
factoring out the strange quark mass and charge, as in Eq. \eqref{eq:2.9}, 
$d_n + d_p =  -  2 g^T_s m_s d_s/3$. 
$g^T_s$ involves sea quarks, and is numerically more challenging to calculate on the lattice than 
$g^T_{u,d}$. 
A recent, state-of-the-art LQCD calculation \cite{Bhattacharya:2015} finds 
$g^T_s = 0.002 \pm 0.011$.
With these large uncertainties, it is not possible to exclude that, due to the 
enhancement $m_s/m_d \sim 20$, the sEDM gives 
a contribution to $d_n + d_p$ of the same size as Eq. \eqref{tensor2}.
The contribution of the sCEDM ${\tilde d}_s$ 
has been addressed in $SU(3)$ $\chi$PT
and QCD sum rules \cite{Pospelov:2005pr,Hisano:2012cc,Fuyuto:2012yf}. 
Refs. \cite{Hisano:2012cc,Fuyuto:2012yf} 
found that ${\tilde d}_s$ could give a large, possibly dominant, contribution 
to the nucleon EDM. 
In the presence of PQ symmetry, ${\tilde d}_s$ does not contribute at LO to 
pion-nucleon $\slashT$ couplings, but affects 
$\bar g_{N\Sigma K}$ 
and thus induces long-range contributions.
The numerical value of the loop is small, but not so small as to compensate 
the enhancement due to the appearance of $m_s$ in the coefficient of the sCEDM.
It would be interesting to test the 
robustness of this prediction by going beyond LO.
In general, the same issue as for $\bar\theta$ remains: the convergence
of $SU(3)$ $\chi$PT.

\section{$\slashT$ Moments of Light Nuclei}
\label{Sec5}

The observation of neutron and proton EDMs will not provide enough 
information to clearly identify the leading $\slashT$ source(s).
$\slashT$  in light nuclei,
being sensitive to different combinations of the couplings in 
Eq. \eqref{eq:3.2}, is highly complementary, and
it can be computed 
in the same theoretical framework as the nucleon EDFF with reliable accuracy.
For these reasons, and with the additional motivation of 
the exciting experimental developments that might allow a direct measurement 
of EDMs of charged ions,
the study of the EDMs of 
deuteron, helion and triton
has received
a lot of attention in the last few years,
both from a phenomenological perspective
\cite{Lebedev:2004va,Liu:2004tq,Stetcu:2008vt,Afnan:2010xd,Song:2012yh}
and with Chiral EFT
\cite{deVries:2011re,deVries:2011an,Bsaisou:2012rg,Bsaisou:2014zwa}.
Very recently, 
the first model calculation of the  $^6$Li EDM has appeared
\cite{Yamanaka:2015qfa}.
For the deuteron, as seen in Tab. \ref{tab1}, two other moments,
MQM and TQM, are sensitive to $\slashT$ and have been calculated 
for $d\le 6$ $\slashT$ sources 
in Refs. \cite{deVries:2011re,Liu:2012tra} and \cite{Mereghetti:2013bta},
respectively, but prospects for their detection are remote at best.
$\slashT$ in neutron-proton \cite{Liu:2006qp}
and neutron-deuteron \cite{Song:2011sw} 
scattering have been revisited recently,
but seem equally difficult to measure.

$\slashT$ is a small effect, and can be treated perturbatively on top 
of the strong nuclear interactions. 
The elements that enter 
EFT calculations in nuclei are
discussed in Section \ref{Sec5.1},
while results for 
nucleon number $A=2,3$ are reviewed
in Section \ref{Sec5.2}. 
The alpha particle ($A=4$) 
has no $\slashT$ moments, but
$\slashT$ FFs of nuclei with a few more nucleons could be calculated 
within this approach.

\subsection{Nuclear Potential and Currents}
\label{Sec5.1}

\begin{figure}
\center
\includegraphics[width = 8cm]{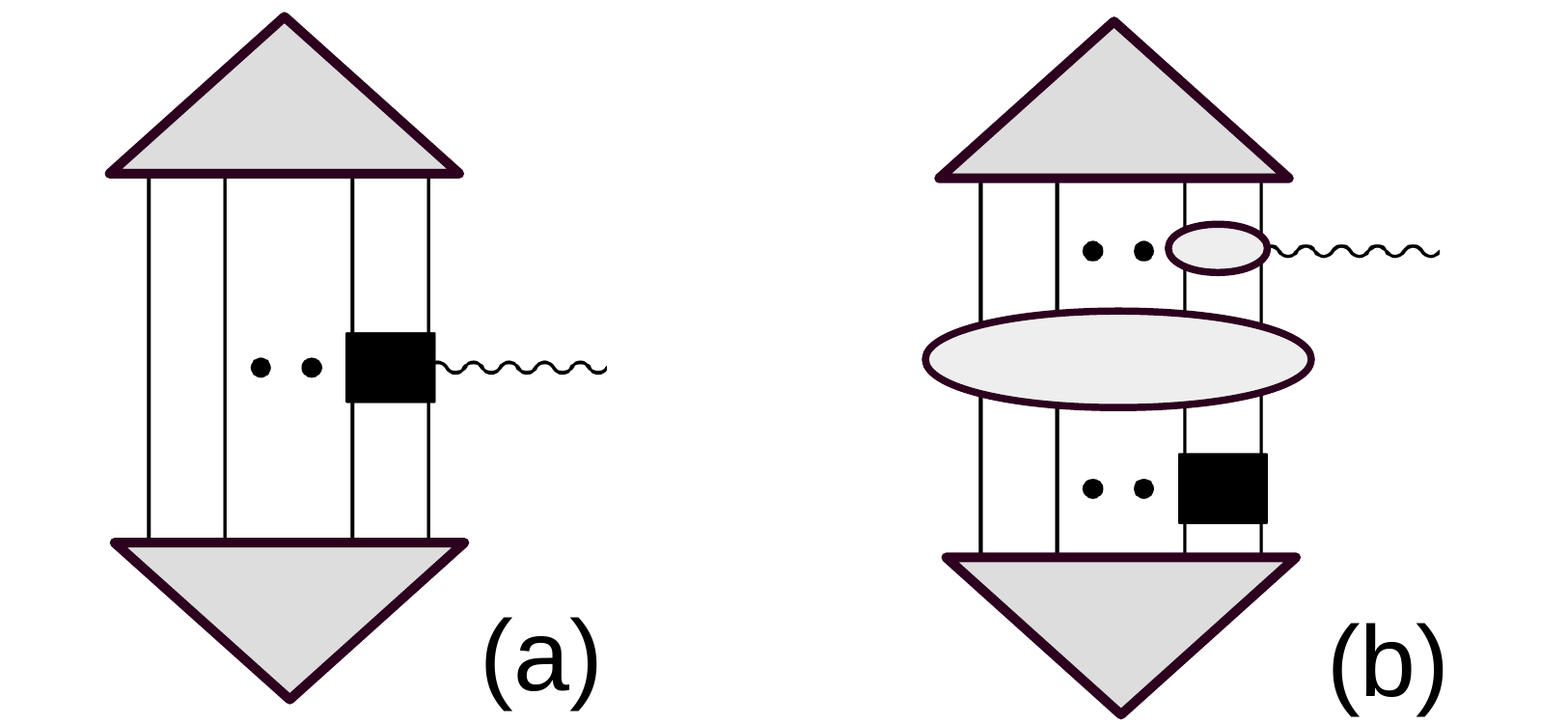}
\caption{Contribution to nuclear EDMs. 
The large triangles denote the nuclear $PT$ wavefunction; 
the oval, iterations of the $PT$ potential; 
the oval with an attached photon, a few-nucleon current;
the black square, the $\slashT$ potential; 
and the large black square with an attached photon, 
a $\slashT$ few-nucleon current. 
Other notation as in Fig. \ref{NEDMLO}.}
\label{Fig4}
\end{figure}

The $\slashT$ electromagnetic current
can be written in the form of Eq. \eqref{eq:4.1}
for a spin-1/2 nucleus, and in a straightforward
generalization that includes MQM \cite{deVries:2011re}
and TQM \cite{Mereghetti:2013bta} for spin-1.
It receives several contributions, schematically 
illustrated in Fig. \ref{Fig4}.
The nuclear wavefunction, denoted by the shaded triangles, as
well as the iteration of the $PT$ potential $V_{PT}$, represented by 
the shaded blob, 
can be obtained by solving the Schr\"odinger equation.
$\slashT$, denoted by black squares,
either affects the 
electromagnetic current 
(Fig. \ref{Fig4}(a)) or perturbs the nuclear wavefunction 
with an insertion of the $\slashT$ potential
(Fig. \ref{Fig4}(b)). 

The power counting of Chiral EFT systematically organizes
the various contributions \cite{deVries:2011an,deVries:2012ab}.
While in the one-nucleon sector a loop brings in a suppression of 
$Q^2/\MQCD^2$,
in problems with two or more nucleons a relative enhancement
of $4\pi m_N/Q$ compensates the loop suppression
in diagrams that involve
intermediate states with nonrelativistic nucleons only, where energies are 
${\cal O}(Q^2/m_N)$ rather than ${\cal O}(Q)$ as in $\chi$PT
\cite{Bedaque:2002mn}.
The one-body contribution in 
Fig. \ref{Fig4}(a) represents the nuclear FF
arising from the EDFF of the constituents, Eq. \eqref{eq:4.2}.
The $PT$ one-nucleon current
contribution to Fig. \ref{Fig4}(b) 
contains an additional loop with respect to 
the one-body contribution in Fig. \ref{Fig4}(a).  
These contributions 
scale as  
\begin{equation}
J_a = \mathcal O\left(F_{0,1}Q\right),
\qquad
J_b  = \mathcal O
\left(\frac{e \bar g_{0,1}}{F^2_{\pi}}Q, e \bar C_{1,2} F^2_{\pi}Q,
\frac{e\bar\Delta}{F^2_{\pi}\MQCD}Q\right).
\label{eq:5.1}
\end{equation}
Depending on the sizes of $\bar g_{0,1}$ compared to $d_{n,p}$, 
insertions of 
two-body currents in Fig. \ref{Fig4}(b)
can be as big as the one-body contribution to
Fig. \ref{Fig4}(a) \cite{deVries:2011an}.

The various ingredients in Fig. \ref{Fig4} can be obtained
explicitly in terms of the $\slashT$ LECs.
For EDMs, the most important $\slashT$ current is the 
sum of one-body contributions,
\begin{equation}
J^0_{\slashTsub,1} (\vec q) = - \frac{i}{2} 
\sum_{i}
\left[ d_n+d_p + \left(d_p-d_n\right) \tau_3^{(i)}\right] \, 
\vec\sigma^{(i)} \cdot \vec q,
\end{equation}
where $\vec\sigma^{(i)}/2$ ($\boldtau^{(i)}/2$)
is the spin (isospin) of the nucleon $i$
that interacts with the photon, and $\vec q$ is the (outgoing)
photon momentum.
Pion-exchange $\slashT$ currents induced by $\bar g_0$, 
which by power counting could be important for the deuteron EDM from 
$\bar\theta$,
are given in Ref. \cite{deVries:2011an},
but their contribution is found to be numerically small.

The tree-level potential induced by the couplings in Eq. \eqref{eq:3.2} 
has a two-body component \cite{Maekawa:2011vs}, which is
generated by one-pion exchange (OPE) with couplings $\bar g_{0,1}$
and by the short-range interactions
$\bar C_{1,2}$, 
\begin{eqnarray}
V_{\slashTsub,2} (\vec k_{i})&=& 
\frac{i}{2F^2_{\pi}} \sum_{j>i}
\vec k_{i}\cdot \left\{
\left[-F^2_{\pi}\bar C_{1} 
+\left(\frac{2g_A\bar g_0}{\vec k_{i}^{\,2} + m^2_{\pi}} 
- F^2_{\pi}\bar C_{2}\right)  \boldtau^{\,(i)} \cdot \boldtau^{\,(j)}\right]
\left(\vec\sigma^{\,(i)} - \vec\sigma^{\,(j)}  \right)  
\right.
\nonumber \\
& &\left.
+ \frac{g_A \bar g_{1}}{\vec k_{i}^{\, 2} + m^2_{\pi}} 
\left[ \left(\tau_3^{(i)} + \tau_3^{(j)}\right)
\left( \vec\sigma^{\, (i)} - \vec\sigma^{\,(j)} \right)  
+ \left(\tau_3^{(i)} -\tau^{(j)}_3 \right) 
\left( \vec \sigma^{(i)} + \vec \sigma^{(j)}\right) 
 \right] 
\right\}, \;\;
\label{eq:5.2.1}
\end{eqnarray}
and a three-body component \cite{deVries:2012ab}, which
is generated by the three-pion coupling $\bar\Delta$,
\begin{eqnarray}
V_{\slashTsub,3}(\vec k_{i}) &=& 
- \frac{2g_A^3 \bar \Delta}{\Fp^4}
\sum_{k>j>i}
\left(\tau_3^{(i)}\,\boldtau^{(j)}\cdot \boldtau^{(k)}
+ \tau_3^{(j)}\,\boldtau^{(i)}\cdot \boldtau^{(k)}
+ \tau_3^{(k)}\,\boldtau^{(i)}\cdot \boldtau^{(j)}\right)
\nonumber\\
&&\times 
\frac{\vec \sigma^{(i)}\cdot \vec k_i \; 
\vec \sigma^{(j)}\cdot \vec k_j \; 
\vec \sigma^{(k)}\cdot \vec k_k}
{(\vec k_i^{\, 2} +\mpi^2)(\vec k_j^{\, 2} +\mpi^2)(\vec k_k^{\, 2} +\mpi^2)}
\label{eq:5.2.2},
 \end{eqnarray}
where $\vec k_{i}$ is the momentum transferred from nucleon $i$.
The 
OPE potential generated by the couplings $\bar g_{0,1}$ is 
well known \cite{Barton:1961eg}. 
It should be important for $\bar\theta$, qCEDM,
gCEDM, PS4QOs, and LR4QOs,
but for the $\bar\theta$ only $\bar g_0$ appears at LO.
The short-range terms, which can be thought as the long-wavelength
effect of $\slashT$ heavy meson exchange,
should be kept at this order for gCEDM and PS4QOs.
The three-body force, which is new, appears at LO only for LR4QOs.
The LO potential induced by the qEDM is given in Ref. \cite{Maekawa:2011vs}, 
but it is of little phenomenological relevance.
Chiral EFT provides a framework to go beyond LO
by including two-pion exchange potentials,
relativistic corrections, 
and OPE potentials from subleading couplings. 
For $\bar\theta$, N$^2$LO corrections to Eq. \eqref{eq:5.2.1} were discussed 
in Ref. \cite{Maekawa:2011vs}. 
The calculation can be immediately extended to the qCEDM, with the observation 
that TPE diagrams involving $\bar g_1$ vanish \cite{Bsaisou:2012rg}.
In the case of the LR4QO, the coupling $\bar\Delta$ gives a 
relatively large NLO correction to the isospin-breaking OPE potential 
\cite{deVries:2012ab,Bsaisou:2014zwa}, the largest
part of which we absorbed in $\bar g_1$ (Eq. \eqref{eq:3.6new}).

\subsection{Moments}
\label{Sec5.2}

The estimate of Eq. \eqref{eq:5.1},
combined with the relative sizes of 
$\bar g_{0,1}$, $\bar d_{0,1}$, $\bar C_{1,2}$ and $\bar\Delta$
discussed in Section \ref{Sec3.2},
lead to the expectation that for qCEDM and LR4QOs
EDMs of light nuclei are dominated by 
the 
$\slashT$ OPE potential.
For the gCEDM and PS4QOs, one-body, 
OPE and $\bar C_{1,2}$ contributions
should be approximately equal, 
while if the qEDM is the only $\slashT$ source
light-nuclear EDMs are well approximated by the 
EDMs of the constituents.
For $\bar\theta$, the situation is more complicated.
For most nuclei, OPE from $\bar g_0$
should dominate,
but when the numbers of protons and neutrons are equal, $N=Z$,
spin/isospin selection rules cause isoscalar $\slashT$ interactions, 
such as $\bar g_0$ and $\bar C_{1,2}$, not to contribute \cite{Haxton:1983dq}
at LO.
$N=Z$ nuclei like the deuteron
are mainly sensitive to isovector
couplings, in particular 
$\bar g_1$, and the EDM from
$\bar\theta$ is suppressed by $Q^2/M_{\textrm{QCD}}^2$
with respect to the naive expectation in Eq. \eqref{eq:5.1}.
Using Eq. \eqref{scalingofdim6}
we obtain the order-of-magnitude estimates 
for 
$d_{d,t,h}$ in Tab. \ref{tab4},
which indicates that light-nuclear EDM measurements would
offer clues regarding $\slashT$ sources.
Expectations about the deuteron MQM are described in Ref. \cite{Liu:2012tra}.

\begin{table}
\tabcolsep5.25pt
\label{tab4}
\begin{center}
\begin{tabular}{c|ccccc}
Source & $\bar\theta$ & qCEDM & qEDM & gCEDM, PS4QOs & LR4QOs 
\tabularnewline \hline 
$\MQCD d_{n}/e$ & 
${\cal O}\left(\frac{m_{\pi}^{2}}{\MQCD^{2}}\bar\theta\right)$ & 
${\cal O}\left(\frac{m_{\pi}^{2}}{M^2_{\slashTsub}}{\tilde{\delta}_i}\right)$ & 
${\cal O}\left(\frac{m_{\pi}^{2}}{M^2_{\slashTsub}}\delta_i\right)$ & 
${\cal O}\left(\frac{\MQCD^{2}}{M^2_{\slashTsub}}(w, \sigma_a)\right)$ &
${\cal O}\left(\frac{\MQCD^{2}}{M^2_{\slashTsub}}\xi\right)$ 
\tabularnewline
$d_{p}/d_{n}$ & ${\cal O}\left(1\right)$ & ${\cal O}\left(1\right)$ & 
${\cal O}\left(1\right)$ & ${\cal O}\left(1\right)$ & ${\cal O}\left(1\right)$
\tabularnewline 
$d_{d}/d_{n}$ & ${\cal O}\left(1\right)$ & 
${\cal O}\left(\frac{\MQCD^{2}}{Q^2}\right)$ &
${\cal O}(1)$ & ${\cal O}(1)$ &
${\cal O}\left(\frac{\MQCD^{2}}{Q^2}\right)$
\tabularnewline 
$d_{h}/d_{n}$ & ${\cal O}\left(\frac{\MQCD^{2}}{Q^2}\right)$ & 
${\cal O}\left(\frac{\MQCD^{2}}{Q^2}\right)$ & ${\cal O}(1)$ & 
${\cal O}(1)$& ${\cal O}\left(\frac{\MQCD^{2}}{Q^2}\right)$
\tabularnewline 
$d_{t}/d_{h}$  & ${\cal O}\left(1\right)$ & ${\cal O}\left(1\right)$ & 
${\cal O}\left(1\right)$ & ${\cal O}\left(1\right)$ & ${\cal O}\left(1\right)$
\tabularnewline \hline
\end{tabular}
\end{center}
\caption{Expected orders of magnitude for the neutron EDM 
(in units of $e/\MQCD$), and for the EDM ratios 
proton to neutron, deuteron to neutron, helion to neutron, 
and triton to helion, 
for $\bar\theta$ and 
$d=6$ sources.  
$Q$ represents the low-energy scales $F_\pi$, $m_\pi$,
and $\sqrt{m_NB}$, with $B$ the binding energy.  (Adapted from Ref.
\cite{deVries:2011an}.)}
\end{table}

In principle, light-nuclear EDMs can be calculated consistently
within EFT. 
Light nuclei are sufficiently diluted to be studied
in an EFT where pions are integrated out \cite{Bedaque:2002mn},
but in this case one cannot easily keep track of the chiral-symmetry
constraints discussed in Section \ref{Sec3}.
For such nuclei, the pions present in Chiral EFT can be treated in 
perturbation theory \cite{Kaplan:1998we}, in which case only contact
interactions need to be included in the $PT$ sector at LO.
For the deuteron, the $PT$ \cite{Kaplan:1998sz} 
and $\slashP T$ \cite{Savage:1999cm} 
electromagnetic FFs have been calculated 
quite successfully in this framework. This framework was extended to
$\slashT$ FFs in Ref. \cite{deVries:2011re},
which gives results similar to an earlier
calculation based on a zero-range model 
\cite{Khriplovich:1999qr},
and in Ref. \cite{Mereghetti:2013bta}.
Similar calculations could be performed for helion and triton. 
Unfortunately, however, pions become non-perturbative at momenta
$Q\sim F_\pi$ \cite{Fleming:1999ee} and thus this approach will fail 
for sufficiently dense nuclei.
Much work has been done with nonperturbative pions following 
Refs. \cite{Weinberg:1990rz,Weinberg:1991um,Ordonez:1992xp},
as reviewed for example in Refs. \cite{Machleidt:2011zz,Epelbaum:2012vx}.
By now, good potentials exist at N$^2$LO 
and N$^3$LO (for example, Ref. \cite{Epelbaum:2004fk}), but 
for only narrow ranges of UV regulators.
Although not consistent from an EFT perspective 
because they lack the necessary counterterms at each
order and generate amplitudes that are not properly renormalized
\cite{Nogga:2005hy},
these potentials produce results in few-body systems that 
are not very different from phenomenological two- plus three-body
potentials 
such as Av18 \cite{Wiringa:1994wb} plus UIX \cite{Pudliner:1997ck}. 
For a calculation of $PT$ FFs for $A=2,3$ systems within this approach,
see Ref. \cite{Piarulli:2012bn}.
Renormalization issues with currents are discussed in
Ref. \cite{Valderrama:2014vra}. 

In Tab. \ref{tab23} we give a sample of
the most recent evaluations of $d_{d,t,h}$
in terms of the nucleon
EDMs, Eqs. \eqref{eq:4.3} and \eqref{eq:4.4},
the pion-nucleon couplings in Eq. \eqref{eq:3.6new}, where we bury some
$\bar\Delta$ contributions,
and the other couplings in Eq. \eqref{eq:3.2}.
For the deuteron, the one-body contribution is given by the isoscalar 
nucleon EDM, $2d_0=d_n+d_p$.\footnote{The deviation of the proportionality
factor from 1 seen in Table \ref{tab23} for the Av18 and N$^2$LO potentials
stems solely from the deuteron $D$-state probability $P_D$.
However, $P_D$ is not an observable \cite{Friar:1979zz},
indicating that the $\sim$ 10\% difference between these
potentials and the perturbative-pion result, where $P_D$ enters only
at N$^2$LO together with other contributions, is within the theoretical error 
of both approaches.} 
Isoscalar $\slashT$ potentials, generated by $\bar g_0$ and $\bar C_{1,2}$, 
vanish on the deuteron. 
$\bar g_0$  does contribute to $d_d$
but only through subleading potentials 
involving $PT$ isospin-breaking couplings, and through two-nucleon 
$\slashT$ currents;
these contributions are small 
\cite{deVries:2011an,Bsaisou:2014zwa}
and omitted here.
The $\bar g_1$ contribution to $d_d$
shows little dependence 
on $V_{PT}$ (see also Ref. \cite{Afnan:2010xd}), 
and it is about a factor of 5 smaller than the expectation 
in Eq. \eqref{eq:5.1}.
$\bar\Delta$ enters only indirectly through $\bar g_1$.

\begin{table}
\tabcolsep5pt
\label{tab23}
\begin{center}
\begin{tabular}{c|c| c c  c c  c c c }            
& Potential (references)
& $d_n$       
& $d_p$  
& $\bar g_0/F_{\pi}$     
& $\bar g_1/F_{\pi}$ 
& $\bar C_1 F^3_{\pi}$
& $\bar C_2 F^3_{\pi}$  
& $\bar \Delta /F_{\pi}m_N$
\\
\hline
& Perturbative pion  \cite{Khriplovich:1999qr,deVries:2011re}
& $1$   
& $1$    
& ---   
& $-0.23$  
& ---  
& ---    
& ---  
\\
 $d_d$
& Av18 \cite{Liu:2004tq,deVries:2011an,Bsaisou:2012rg,Bsaisou:2014zwa,Yamanaka:2015qfa}
& $0.91$  
& $0.91$  
& ---  
& $-0.19 $    
& ---  
& ---    
& ---  
\\
& N$^2$LO  \cite{Bsaisou:2012rg,Bsaisou:2014zwa}
& $0.94$    
& $0.94$    
& ---  
& $-0.18$    
& ---  
& ---    
& ---  
\\
\hline
& Av18 \cite{Stetcu:2008vt,deVries:2011an,Yamanaka:2015qfa}     
& $-0.05 $   
& $0.90$    
& $0.15$  
& $-0.28$ 
& $0.01$ 
& $-0.02$    
& n/a  
\\
 $d_t$
& Av18+UIX \cite{Song:2012yh,Bsaisou:2014zwa}     
& $-0.05 $   
& $0.90$    
& $0.07$  
& $-0.14$ 
& $0.002$ 
& $-0.005$   
& $0.02$
\\
& N$^2$LO \cite{Bsaisou:2014zwa}   
& $-0.03$    
& $0.92$    
& $0.11$   
& $-0.14$ 
& $0.05$ 
& $-0.10$ 
& $0.02$ 
\\
\hline
& Av18 \cite{Stetcu:2008vt,deVries:2011an,Yamanaka:2015qfa}
& $0.88$    
& $-0.05$   
& $-0.15$  
& $-0.28$ 
& $-0.01$ 
& $0.02$   
& n/a
\\
 $d_h$
& Av18+UIX \cite{Song:2012yh,Bsaisou:2014zwa}     
& $0.88$    
& $-0.05$   
& $-0.07$  
& $-0.14$ 
& $-0.002$ 
& $0.005$   
& $0.02$
\\
& N$^2$LO \cite{Bsaisou:2014zwa}   
& $0.90$    
& $-0.03$    
& $-0.11$   
& $-0.14$ 
& $-0.05$ 
& $0.11$ 
& $0.02$ 
\\
\hline
\end{tabular}
\caption{Dependence of 
the deuteron, triton and helion EDMs 
on $\slashT$ LECs
for various $PT$ potentials.
Entries are dimensionless in the first two columns
and in units of $e\, \textrm{fm}$ in the remaining columns.
``---'' indicates very small numbers.}
\end{center}
\end{table}

The situation is strikingly different for helion and triton.
The one-body contributions are given mostly
by, respectively, $d_n$ and $d_p$.\footnote{In analogy to the isoscalar trinucleon MDM 
\cite{Friar:1979zz}, one expects model-dependent contributions from the 
trinucleon $D$-state probability to $d_t+d_h$ of similar size as
those from $P_D$ to $d_d$, see previous footnote.}
$\bar g_0$ and $\bar g_1$ contribute 
at about the same level. In particular, $\bar g_1$
contributes to the isoscalar combination 
$d_t + d_h$, while $\bar g_0$ to the isovector, $d_t-d_h$.
There is a factor-of-2 disagreement between
calculations based on the No-Core Shell Model 
\cite{Stetcu:2008vt,deVries:2011an}
and on the Faddeev equation \cite{Song:2012yh,Bsaisou:2014zwa}.
As for the deuteron, the EDM induced by the OPE $\slashT$ potential 
is a few times smaller than the expectation in Eq. \eqref{eq:5.1}.
Changing 
$V_{PT}$ has little effect on the contribution of $\bar g_1$, 
while $\bar g_0$ is more affected.
The isoscalar 
couplings $\bar C_{1,2}$ give 
a nonvanishing contribution to 
$d_t - d_h$.
These operators are the most sensitive to the choice of $V_{PT}$ 
and to the details of the nuclear calculation, 
like the choice of regulator \cite{deVries:2011an,Bsaisou:2014zwa},
and more conclusive results have to wait for a more consistent approach.
Still, also this contribution 
appears to be smaller than predicted by Eq. \eqref{eq:5.1}.
$\bar\Delta$ now enters explicitly through
the three-nucleon $\slashT$ force, Eq. \eqref{eq:5.2.2},
but its contribution is smaller than expected by power counting
and can be neglected \cite{Bsaisou:2014zwa}.

Table \ref{tab23}
allows us to draw more precise
conclusions about the relative size of EDMs,
which also qualify some of the expectations in Tab. \ref{tab4}.
The expectation of relatively large light-nuclear
EDMs for $\bar\theta$, qCEDM and LR4QOs
is tampered by the small effects of the OPE
$\slashT$ potential shown in Tab. \ref{tab23}.
Still, the even smaller effects from the rest of
the $\slashT$ potential leave some opportunities open:

\begin{itemize}

\item
$\bar\theta$:
NDA suggests that $\bar C_{1,2}$
and explicit $\bar\Delta$ effects can be neglected.
Using the estimates for $\bar g_{1}$ 
in Eq. \eqref{eq:3.6} 
and the LQCD evaluation of $d_{n,p}$ in
Eq. \eqref{latticenuc}, 
\begin{equation}
d_d
\simeq 0.9 ( d_n + d_p ) - 0.2 \frac{\bar g_1}{F_{\pi}} ,
\qquad
\frac{d_t+ d_h}{2} 
\simeq   0.9 \frac{d_n + d_p}{2} - 0.1 \frac{\bar g_1}{F_{\pi}},
\label{eq:5.4}
\end{equation}
which satisfy, within large uncertainties,
the NDA expectation that 
the deuteron EDM and the helion/triton isoscalar EDM combination 
receive contributions of similar size from the isoscalar 
nucleon EDM and $\bar g_1$.
When $d_n + d_p$ is known, from either experiment or a more precise LQCD extraction, 
a measurement of either $d_d$ or $d_t + d_h$ 
will allow a determination of $\bar g_1$.
A value of $\bar g_1$ much larger than the estimate in Eq. \eqref{eq:3.6}
or a 
$d_d$ much larger than $d_n$ or $d_p$ 
will point to $\slashT$ of non-$\bar\theta$ origin.
The helion/triton isovector combination,
\begin{equation}
\frac{d_t - d_h}{2} 
\simeq  0.9 \frac{d_p - d_n}{2} - 0.1 \frac{\bar g_0}{F_{\pi}} ,
\label{eq:5.5}
\end{equation}
is expected to be dominated by OPE,
but the smallness of $\bar g_0$ and the relative suppression of 
the OPE contribution conspire      
to enhance the importance of the one-body contribution.  
Nonetheless, once $d_p - d_n$ is measured,  Eq. \eqref{eq:5.5} 
allows the extraction of $\bar g_0$, and, through Eq. \eqref{eq:3.6zero}, 
the determination of $\bar\theta$, without further nonperturbative input.

\item
qCEDM: As for $\bar\theta$,
$\bar C_{1,2}$ and $\bar\Delta$ should be higher order, contributing 
at N$^2$LO and N$^3$LO respectively, so Eqs. \eqref{eq:5.4} and \eqref{eq:5.5}
also hold.
In this case, however, the couplings are not well determined.
If we assume that $d_{0,1}$
are approximated by their non-analytic pieces 
and that $\bar g_0 \sim \bar g_1$, 
then $d_d$ and  $d_t + d_h$  
would be dominated by $\bar g_1$, with $d_0$ contributing at the $10\%$ level. 
Due to the logarithmic enhancement in $d_p - d_n$
and the relative suppression of OPE in $d_{t,h},$
the one-body and OPE contributions to $d_t - d_h$
should be roughly of the same size. 
Obviously, this is not a firm conclusion
and a better grip on $d_{n,p}$ and $\bar g_{0,1}$ is needed from LQCD.

\item
LR4QOs: Similar considerations hold,
except that the coupling $\bar g_0$ should be small 
and only $d_d$ and  $d_t + d_h$  
could be enhanced with respect to $d_{n,p}$.

\item
gCEDM and PS4QOs:
The short-range contributions from $\bar C_{1,2}$ 
are expected to be as important as the one-body contribution and OPE,
but for the systems considered here the 
effects of the $\slashT$ potential are numerically smaller than expected,
although highly dependent on the short-distance treatment of $V_{PT}$.
If this result stands, Eqs. \eqref{eq:5.4} and \eqref{eq:5.5} hold again
and
also in this case LQCD input is sorely needed.

\item
qEDM: the simplest and most predictive situation, in which all EDMs 
should be well approximated by one-body contributions, and 
both $\bar g_{0,1}$
can be dropped in Eqs. \eqref{eq:5.4} and \eqref{eq:5.5}.

\end{itemize}

Our discussion has underlined how the 
EDMs of light nuclei 
complement the nucleon EDM,
and can play an important role in singling out the 
microscopic source of $\slashT$
---for specific BSM examples, see Ref. \cite{Dekens:2014jka}.
Qualitatively, EDMs of nuclei with $N=Z$ are particularly interesting:
an observation in these systems of a large EDM, compared to the nucleon EDM, 
would suggest that $\slashT$
does not come from $\bar\theta$, 
but rather from 
isospin-breaking sources, like qCEDM or LR4QOs.
Quantitative conclusions require more precise evaluations of the couplings 
in Eq. \eqref{eq:3.2}, a program  
under way using LQCD.

\section{Outlook}
\label{Sec6}

We reviewed here the main steps that link $\slashT$ in the SM and beyond
to observables where dramatic experimental progress is expected, 
nucleon and light-nuclear EDMs. 
The key idea is that 
EFTs enable us to keep track of symmetries across scales, 
with minimal assumptions about unknown dynamics.
As a consequence, the various spin, isospin, and spatial profiles
of light nuclei probe different aspects of (B)SM $\slashT$,
and various (admittedly difficult) 
precision measurements would allow precious information 
on physics at a scale comparable, or perhaps beyond, what can be reached
in the energy frontier.

We sketched this link in broad terms, with an emphasis on progress
over the last five-ten years. Although the framework is in place,
we emphasized a few of the aspects where gaps remain.
Perhaps the most pressing are
{\it i)} a more systematic assessment of 
the effects of flavor-changing operators in the running
of the more relevant light-quark operators;
{\it ii)} LQCD calculations of the LECs
of the $\slashT$ Lagrangian, either directly or making use
of chiral symmetry to relate $\slashT$ matrix elements to
$T$ quantities;
and 
{\it iii)} a fully consistent EFT treatment of the $PT$ dynamics
in nuclei beyond the deuteron.

The usefulness of this framework does not hinge solely
on experiments on light nuclei. Atomic and molecular EDMs 
(in particular, from diamagnetic atoms) also
depend to some extent on nuclear $\slashT$ (especially
Schiff moments), albeit for
much heavier nuclei than we have discussed here.
From the nuclear physics perspective, here lies the most
important challenge: to extend the EFT approach beyond light nuclei.
The last decade has witnessed extraordinary progress in the development
of the so-called {\it ab initio} methods, which aim to calculate
the properties of medium-mass nuclei starting from given internucleon
forces.
Most of this work, just like the trinucleon calculations presented here,
treats the input, whether inspired by EFT
or not, as a phenomenological
potential where there is no hierarchy of interactions.
With such a hybrid approach, the $\slashT$ properties 
of other nuclei could certainly
be calculated using as input not only a $PT$ potential inspired by
EFT, but also the $\slashT$ interactions in Eq. \eqref{eq:3.2}.
This would be an alternative to existing calculations
(reviewed, for example, in Ref. \cite{Engel:2013lsa}) which are typically
based on three nonderivative pion-nucleon couplings
and uncontrolled (but perhaps valid) approximations,
such as random-phase or mean-field. 
We hope this review serves to stimulate such alternative calculations.

On a longer time scale, a fully consistent EFT formulation for larger nuclei 
would be desirable. {}From an {\it ab initio} perspective, nuclear EFT
interactions cannot be treated as a black box, but instead subleading
interactions should be included in perturbation theory to avoid
extraneous regulator dependence. {}From a more effective perspective,
an in-medium EFT appropriate for heavy nuclei should be formulated.
The many regularities found among nuclear properties offer 
great opportunities for the effective field theorist.

\section*{Acknowledgments}
Our understanding of 
$\slashT$ benefited tremendously
from the insights of our collaborators 
V. Cirigliano, J. de Vries, J. Engel, R. Higa, W. Hockings,
C.-P. Liu, C. Maekawa, M. Ramsey-Musolf, I. Stetcu, R. Timmermans
and A. Walker-Loud.
We thank J. de Vries for comments on the manuscript, and
T. Blum, T. Izubuchi, E. Shintani and B. Yoon for providing us with 
LQCD results 
and for interesting discussions. 
The research of EM was supported by the LDRD program at 
Los Alamos National Laboratory.
This material is based upon work supported 
in part by the U.S. Department of Energy, 
Office of Science, Office of Nuclear Physics, 
under Award Number DE-FG02-04ER41338. 

\bibliography{AReview1}

\bibliographystyle{ar-style5}

\end{document}